\documentclass[journal,romanappendices]{IEEEtran}

\makeatletter
\def\ps@headings{%
\def\@oddhead{\mbox{}\scriptsize\rightmark \hfil \thepage}%
\def\@evenhead{\scriptsize\thepage \hfil\leftmark\mbox{}}%
\def\@oddfoot{}%
\def\@evenfoot{}}
\makeatother
\usepackage{latexsym}
\usepackage{amssymb}
\usepackage{epsfig}
\usepackage{amsthm} 
\usepackage{subfigure}
\usepackage{graphics,color}
\usepackage{graphicx}
\usepackage{amsmath}
\usepackage[ruled,vlined,boxed,linesnumbered]{algorithm2e}
\usepackage{cite}
\usepackage{comment}
\usepackage{url}
\usepackage{xspace}
\usepackage{psfrag}
\usepackage{booktabs}
\usepackage{stfloats}
\usepackage{balance}
\usepackage{soul}

\newtheorem{theorem}{Theorem}
\newtheorem{lemma}{Lemma}
\newtheorem{proposition}{Proposition}
\newtheorem{corollary}{Corollary}

\newtheorem{fact}{Fact}
\theoremstyle{definition}

\newtheorem{remarks}{Remark}

\newtheorem{claim}{Claim}

\def\fc{{\mathfrak c}}

\def\Pe{\mathrm {Pe}}
\def\Pr{\mathrm {Pr}}
\def\tmin{\mathrm {min}}

\def\taue{{\tau_\epsilon}}
\def\tause{{\tau^*_\epsilon}}
\def\taute{{\tilde\tau_\epsilon}}
\def\tauti{{\tilde\tau_\iota^*}}

 \DeclareMathOperator*{\argmax}{arg\,max}
 \DeclareMathOperator*{\argmin}{arg\,min}

\newcommand{\ignore}[1]{}

\begin{document}

\title{Extrinsic Jensen--Shannon Divergence: \\ Applications to Variable-Length Coding}
\author{Mohammad Naghshvar, Tara Javidi, and Mich\`ele Wigger 
\thanks{This paper was presented in part at ITW 2012 and ISITA 2012. 

M. Naghshvar was with the Department of Electrical and Computer Engineering,
University of California San Diego, La Jolla, CA 92093 USA. He is now with
Qualcomm Technologies Inc., San Diego, CA 92121 USA (e-mail: mnaghshv@qti.qualcomm.com).
T. Javidi is with the Department of Electrical and Computer
Engineering, University of California San Diego, La Jolla, CA 92093 USA
(e-mail: tjavidi@ucsd.edu).
M. Wigger is with the Department of Communications and Electronics, Telecom ParisTech, Paris, France. 
(e-mail: michele.wigger@telecom-paristech.fr).


The work of M.~Naghshvar and T.~Javidi was partially supported by the industrial sponsors of UCSD Center for Wireless Communication (CWC) 
and NSF Grants CCF-1018722 and CCF-1302588. 
The work of M.~Wigger was partially supported by the city of Paris under the program ``Emergences''.

Copyright (c) 2014 IEEE. Personal use of this material is permitted.  However, permission to use this material for any other purposes must be obtained from the IEEE by sending a request to pubs-permissions@ieee.org.
%
%
%
%
}}

\maketitle

\thispagestyle{empty}

\begin{abstract}
This paper considers the problem of variable-length coding over a discrete memoryless channel (DMC) with noiseless feedback. The paper provides a stochastic control view of the problem whose solution is analyzed via a newly proposed symmetrized divergence, termed extrinsic Jensen--Shannon (EJS) divergence. It is shown that strictly positive lower bounds on EJS divergence provide non-asymptotic upper bounds on the expected code length. The paper presents strictly positive lower bounds on EJS divergence, and hence non-asymptotic upper bounds on the expected code length,  for the following two coding schemes: variable-length posterior matching and MaxEJS coding scheme which is based on a greedy maximization of the EJS divergence. 

As an asymptotic corollary of the main results, this paper also provides a rate--reliability test. Variable-length coding schemes that satisfy the condition(s) of the test for parameters $R$ and $E$, are guaranteed to achieve rate $R$ and error exponent $E$. The results are specialized for posterior matching and MaxEJS to obtain deterministic one-phase coding schemes achieving capacity and optimal error exponent. For the special case of symmetric binary-input channels, simpler deterministic schemes of optimal performance are proposed and analyzed.    
%
%
\end{abstract}


\section{Introduction}

In his seminal paper \cite{Burnashev76}, Burnashev provided upper and lower bounds on the minimum expected number of channel uses $\mathbb{E}[\tause]$ that are needed to convey a message (from a fixed message set of size $M$) 
with average probability of error smaller than some~$\epsilon$ over a discrete memoryless channel (DMC) with noiseless feedback. 
For all code rates below the capacity of the DMC, the ratio between the upper and lower bounds 
approaches 1 as $\epsilon \to 0$.
Therefore, the bounds yield the optimal error exponent, also referred to as Burnashev's reliability function
 \begin{align}
 \label{OptExp}
 E(R) := \lim_{\epsilon \to 0} \frac{-\log \epsilon}{\mathbb{E}[\tause]}=C_1\left(1-\frac{R}{C}\right)
 \end{align} where $C$ denotes the capacity of the channel, $R\in [0,C]$ is the expected rate of the code, 
and $C_1$ is the maximum Kullback--Leibler (KL) divergence between the conditional output distributions given any two inputs. 

Burnashev proved the upper bound using a two-phase coding scheme.
In the first phase, referred to as the \emph{communication} phase, the
transmitter tries to increase the decoder's belief about the true message.
At the end of this phase, the message with the highest posterior probability
is selected as a candidate. 
The second phase, referred to as the \emph{confirmation} phase,
serves to verify the correctness of the output of phase one.
Subsequently, in \cite{Yamamoto79,Ooi98}   
alternative two-phase coding schemes attaining Burnashev's reliability function were provided{, while it was shown in
\cite{Caire06} that Burnashev's communication phase can be replaced with \emph{any} capacity achieving block code.}
In \cite{Tchamkerten06}, Burnashev's reliability function was shown to be attainable
using a two-phase scheme for a binary symmetric channel (BSC) with an unknown crossover probability.  
In \cite{Nakiboglu08}, Burnashev's reliability function was extended to the cost constrained case, and the achievability was proved via a two-phase coding scheme generalizing that of \cite{Yamamoto79}.

In \cite{Horstein63, Burnashev74}, see also \cite{Shayevitz11}, a one-phase scheme
for transmission over a BSC with noiseless feedback was proposed.
This scheme, first proposed in \cite{Horstein63}, is briefly explained next.
Each message is represented as a subinterval of size $\frac{1}{M}$ of the unit interval. 
After each transmission and given the channel output,
the posterior probability of all subintervals are updated.
In the next time slot, the transmitter sends~0
if the true message's corresponding subinterval is below the current median, 
or~1 if it is above. 
If the current median lies within the true message's subinterval,
then the transmitter sends 0 and 1 randomly according to weights determined 
by the length of the portions of the subinterval above and below the median.
As the rounds of transmission proceed, the posterior 
probability of the true message's subinterval most likely grows larger than $\frac{1}{2}$, which pushes the median within the message's subinterval and thus leads to a randomized encoding.
In a fixed-length setting, this simple one-phase scheme is known to achieve the capacity of a BSC \cite{Burnashev74}, and its posterior matching extension has recently been shown to achieve the capacity of general DMCs with noiseless feedback \cite{Shayevitz11}. 
Li and El Gamal \cite{LiGamal14} proposed a variant of the posterior matching scheme and derived a lower bound on its error exponent in the fixed-length setting.

These previous results raise the question whether having two separate phases of operation and 
randomized encoding are necessary to achieve Burnashev's reliability function or not. 
In this paper we show that this is not the case in the variable-length setting. 
In particular, we propose a deterministic one-phase\footnote{This means that there exists a stationary encoding strategy which performs roles of \emph{communication} and \emph{confirmation} when necessary.} coding scheme which is proved to achieve Burnashev's reliability function of the DMC with noiseless feedback.

More generally, the main contributions of the paper are:
 
 \begin{itemize}
 \item 
Drawing parallels between mutual information and symmetrized L~divergence \cite{Lin91}, the \emph{extrinsic Jensen--Shannon (EJS) divergence} of the conditional output distributions 
	with respect to the receiver's posterior probability is proposed 
	as the key performance measure of any given coding scheme. 

	\item The main result is to show that strictly positive lower bounds on the EJS divergence provide a \emph{non-asymptotic} 
		upper bound on the expected number of channel uses necessary for a coding scheme to 
		obtain a given (arbitrarily small) error probability. 
 
		\item As a corollary, a rate--reliability test for variable-length 
		coding schemes is proposed. That means, lower bounds on the the EJS divergence immediately convert to lower bounds on the rates and  error exponents achieved by a given coding scheme.
		
		\item	The test is  utilized to show that MaxEJS, a newly proposed one-phase coding scheme  that maximizes EJS divergence in each step, achieves the optimal error exponent of the DMC with noiseless feedback in the variable-length setting.
		
		\item The test is also utilized to 
		provide an alternative (simple and concise) proof that the variable-length version of posterior 
		matching achieves capacity when $C_1<\infty$. 
		Furthermore, an achievable error exponent is obtained for variable-length posterior matching.  	
	
\end{itemize}

The proof of the main result---lower bounds on EJS divergence provide a non-asymptotic upper bound on the expected number of channel uses required for a given probability of error---is very succinct and follows a new technique as described below:
\begin{itemize}
	\item This paper provides a stochastic control view of the problem of variable-length coding with feedback. 
	This stochastic 
	control problem, a discrete version of that suggested in \cite{toddISIT}, is analyzed via a Lyapunov type argument for Markov decision problems.
	
	\item It is shown that an appropriate (Lyapunov type) functional, closely related to average log-likelihood, of the posterior 
	is a submartingale whose expected drift can be expressed in terms of EJS symmetrized divergence. 
	\item The level crossing stopping time associated with a submartingale is shown to be upper bounded 
	via a lower bound on the EJS divergence obtained at each stage of encoding. 
	
\end{itemize}

The remainder of this paper is organized as follows. 
In Section~\ref{Sec:EJS}, we introduce the EJS divergence
and discuss some of its properties.
In Section~\ref{Sec:VLcoding}, we formulate the problem of channel coding with noiseless feedback.
Section~\ref{Sec:Main} provides the main results of the paper for general DMCs: i) an EJS divergence based non-asymptotic analysis of variable-length coding, ii) a specialization of this analysis to variable-length posterior matching, and iii) a specialization to a new deterministic one-phase coding scheme that is based on greedy maximization of the EJS divergence. 
In Section~\ref{Sec:Sym}, we consider the special case of symmetric binary-input channels 
and propose simple deterministic schemes.  
Finally, in Section~\ref{sec:rate-reliability}, we analyze the achievable rates and error exponents of the coding schemes presented in the previous two sections.

We finish this section with some notation.

\underline{Notation}:
Let $[x]^+=\max \{x,0\}$. 
The indicator function $\mathbf{1}_{\{A\}}$ takes the value 1 whenever event $A$ occurs, and 0 otherwise.
The $i^\mathrm{th}$ element of vector $\boldsymbol{v}$ is denoted by $v_{i}$. 
For any set $\mathcal{S}$, $\left| \mathcal{S} \right|$ denotes the cardinality of $\mathcal{S}$.
%
All logarithms are in base 2.
The entropy function on a vector $\boldsymbol{\rho}=[\rho_1,\rho_2,\ldots,\rho_M] \in [0,1]^M$
is defined as $H(\boldsymbol{\rho}):=\sum_{i=1}^{M} \rho_i \log\frac{1}{\rho_i}$, with the convention that $0 \log\frac{1}{0} = 0$. 
%
%
We denote the conditional probability $P(Y|X=x)$ by $P_x$. 


\section{Preliminaries}
\label{Sec:EJS}

\subsection{Known Symmetric Divergences and Mutual Information}

We first recall some well known divergences. The \emph{Kullback--Leibler (KL) divergence} between two probability distributions $P_Y$ and $P_Y'$ over a finite set $\mathcal{Y}$ is defined as 
$D(P_{Y}\|P_Y'):=\sum_{y \in \mathcal{Y}} P_Y(y) \log\frac{P_Y(y)}{P_Y'(y)}$ with the convention $0 \log \frac{a}{0}=0$ and $b \log \frac{b}{0}=\infty$ for $a,b\in [0,1]$ with $b\neq 0$. 
The KL divergence satisfies the following lemma.
\begin{lemma}
\label{DPQa}
For any two distributions $P$ and $Q$ on a set $\mathcal{Y}$ and $\alpha\in[0,1]$, 
$D(P\|\alpha P + (1-{\alpha}) Q)$ is decreasing in $\alpha$.
\end{lemma}
\begin{IEEEproof}
Let  $\beta\in[0,1]$ satisfy $\beta \le \alpha$. Then,
\[
\alpha P + (1-\alpha) Q = \gamma \left(\beta P + (1-\beta)Q \right) + (1-\gamma) P
\]
 where $\gamma = \frac{1-{\alpha}}{1-{\beta}} \le 1$. By Jensen's inequality and the convexity of the KL~divergence:
 \begin{IEEEeqnarray}{rCl}\lefteqn{
D\big(P \| \alpha P+(1-{\alpha} )Q\big)} \nonumber \qquad \\
 & \leq & \gamma D\big(P \| \beta P + (1-{\beta}) Q\big) + (1-\gamma) D\big(P\| P\big) \nonumber\\
& \leq &  D\big(P \| \beta P + (1-\beta) Q\big)
 \end{IEEEeqnarray}
 where the last inequality follows because $D\big(P\|P\big)=0$ and $\gamma \leq 1$.
\end{IEEEproof}
The KL divergence is \emph{not} symmetric, i.e., in general $D(P_Y\|P_Y') \neq D(P_Y'\|P_Y)$.
The \emph{J~divergence} \cite{Jeffreys46} and \emph{L~divergence} \cite{Lin91} symmetrize the KL divergence: 
\begin{align}
J(P_1,P_2) &:= D(P_1\|P_2) + D(P_2\|P_1),\\
L(P_1,P_2) &:= D\Big(P_1\|\frac{1}{2}P_1+\frac{1}{2}P_2\Big) + D\Big(P_2\|\frac{1}{2}P_1+\frac{1}{2}P_2\Big).
\end{align} 
The L~divergence can also be related to the \emph{Jensen difference} with respect to the Shannon entropy function~\cite{Burbea82b}: 
\begin{align}
\label{L-H}
{\frac{1}{2}}L(P_1,P_2) &= H\left(\frac{1}{2}P_1+\frac{1}{2}P_2\right) - \left(\frac{1}{2}H(P_1) + \frac{1}{2}H(P_2)\right),
\end{align} 
where for $P$ a probability mass function over $\mathcal{X}$, we have $H(P) := -\sum_{x\in\mathcal{X}} P(x)\log P(x)$.
Let $\Theta$ be a random variable that uniformly takes values in $\{1,2\}$ and $Y \sim P_{\Theta}$ 
(which implies that $\Pr(Y=y)=\frac{1}{2} P_{1}(y) + \frac{1}{2} P_{2}(y)$). From~\eqref{L-H},
\begin{align}
\label{L-I}
{\frac{1}{2}}L(P_1,P_2) = H(Y)-H(Y|\Theta)=I(\Theta;Y)
\end{align}
where $H(Y):=H(P_{\Theta})$  is the entropy of $Y$ and $H(Y|\Theta):= \sum_{\theta=1,2} \frac{1}{2}H(P_{\theta})$ 
the conditional entropy of $Y$ given $\Theta$;  
$I(\Theta;Y)$ is called the \emph{mutual information} between $\Theta$ and $Y$.

The \emph{Jensen--Shannon (JS) divergence} \cite{Burbea82b,Lin91} is defined similarly to the L~divergence but for general $M\geq 2$ probability distributions. 
Given $M$ probability distributions $P_1, P_2 \ldots, P_M$ over a set $\mathcal{Y}$ and 
a vector of a priori weights $\boldsymbol{\rho}=[\rho_1,\rho_2,\ldots,\rho_M]$,
where $\boldsymbol{\rho}\in[0,1]^M$ and $\sum_{i=1}^M \rho_i=1$, the JS divergence is defined as
\cite{Burbea82b,Lin91}:
\begin{align}
\label{JS-H}
\nonumber
JS(\boldsymbol{\rho};P_1,\ldots,P_M) &:= \sum_{i=1}^M \rho_i D\bigg(P_i\|\sum_{j=1}^M \rho_j P_j\bigg)\\
&= H\bigg(\sum_{i=1}^M \rho_i P_i\bigg) - \sum_{i=1}^M \rho_i H(P_i).
\end{align}
Let $\Theta$ be a random variable that takes values in $\{1,2,\ldots,M\}$ and has probability mass function $\boldsymbol{\rho}$ and $Y \sim P_{\Theta}$ (which implies that $\Pr(Y=y)=\sum_{i=1}^M \rho_i P_{i}(y)$). From~\eqref{JS-H},
\begin{align}
\label{JS-I}
JS(\boldsymbol{\rho};P_1,\ldots,P_M)= H(Y)-H(Y|\Theta)=I(\Theta;Y).
\end{align}

\subsection{A New Divergence: Extrinsic Jensen--Shannon Divergence}
We introduce the \emph{extrinsic Jensen--Shannon (EJS) divergence} which extends the J divergence  for general $M\geq 2$ probability distributions $P_1, P_2, \ldots, P_M$ and for an $M$-dimensional weight vector~$\boldsymbol{\rho}$:
\begin{subequations}\label{eq:EJSdef}
\begin{align}
EJS(\boldsymbol{\rho};P_1,\ldots,P_M) := \sum_{i=1}^M \rho_i D\bigg(P_i\|\sum_{j\neq i} \frac{\rho_j}{1-\rho_i} P_j\bigg)
\end{align}
when $\rho_i<1$ for all $i\in\{1,\ldots, M\}$, and as
\begin{align}
EJS(\boldsymbol{\rho};P_1,\ldots,P_M) := \max_{j\neq i} D(P_i\|P_j)
\end{align} 
\end{subequations}
when $\rho_i=1$ for some $i\in\{1,\ldots,M\}$.


Let $U(\cdot)$ denote the average log-likelihood function:
\begin{equation}\label{def:U}
U(\boldsymbol{\rho}):=\sum_{i=1}^{M} \rho_i \log\frac{1-\rho_i}{\rho_i}.
\end{equation} 

\begin{lemma}[Properties of EJS Divergence]\label{lem:propEJS}
The EJS divergence $EJS(\boldsymbol{\rho};P_1,\ldots, P_M)$ as defined in~\eqref{eq:EJSdef} satisfies the following three properties. 
\begin{enumerate}
\item It is lower bounded by the JS divergence: \label{item:EJS-JS}
\begin{equation}\label{EJS-JS}
EJS(\boldsymbol{\rho};P_1,\ldots,P_M) \ge JS(\boldsymbol{\rho};P_1,\ldots,P_M). 
\end{equation}
\item It can be expressed as \label{item:EJS-U}
\begin{align}\label{EJS-U}
\lefteqn{EJS(\boldsymbol{\rho};P_1,\ldots,P_M)} \\ 
&= U(\boldsymbol{\rho}) - \sum_{y\in\mathcal{Y}} P_{\boldsymbol{\rho}}(y) U\Big(\Big[\frac{\rho_1 P_1(y)}{P_{\boldsymbol{\rho}}(y)},\ldots,\frac{\rho_M P_M(y)}{P_{\boldsymbol{\rho}}(y)}\Big]\Big) \nonumber
\end{align}
where $P_{\boldsymbol{\rho}}(y)=\sum_{i=1}^M \rho_i P_{i}(y)$.
\item \label{item:EJS-cnvx}
It is convex in the distributions $P_1,\ldots, P_M$.
\end{enumerate}
\end{lemma}
The proof of Lemma~\ref{lem:propEJS} is given in Appendix~\ref{app:EJSproperties}.

Equation~\eqref{JS-H} shows that if the entropy function $H(\cdot)$ is used to measure uncertainty, then the expected reduction in uncertainty can be characterized by the JS divergence (or equivalently, the mutual information).
Similarly, Equation~\eqref{EJS-U} implies that the EJS divergence characterizes the expected reduction in uncertainty when uncertainty is measured via the average log-likelihood function $U(\cdot)$. This will be a key point when we derive our main results for the problem of variable-length coding with feedback. In fact we analyze the performance of different coding schemes by their expected reduction in uncertainty, measured by EJS divergence, after every transmission.

\begin{remarks}
The EJS divergence defined in this paper is not the unique generalization of the J~divergence.
There exist other $M$-dimensional generalizations of the J~divergence such as 
$\sum_{i=1}^M \rho_i \sum_{j=1}^M \rho_j J(P_i,P_j)$ which was studied in~\cite{Toussaint71}.
However, as will be discussed in details later in the paper, properties of EJS such as the one provided by \eqref{EJS-U} above makes it a {suitable} measure of information for our applications of interest.  
\end{remarks}

\begin{remarks}
Given a uniform prior, the \emph{full anthropic correction} proposed in the context of mutual information estimation~\cite{Gastpar2010} is a special case of the EJS divergence between the corresponding empirical distributions obtained via sampling. In particular, the authors in~\cite{Gastpar2010} used the notion of anthropic correction as an estimator of the mutual information between signals acquired in neurophysiological experiments where only a small number of stimuli can be tested. 
\end{remarks}

\section{Coding over DMC with Noiseless Feedback}
\label{Sec:VLcoding}

\subsection{The Problem Setup}

\begin{figure}[htp]
\centering
\psfrag{t}{{$\Theta$}}
\psfrag{x}{\hspace*{-0.05in}{$X_t$}}
\psfrag{y}{{$Y_t$}}
\psfrag{z}{{$Y^{t-1}$}}
\psfrag{h}{{$\hat{\Theta}$}}
\includegraphics[width=0.489\textwidth]{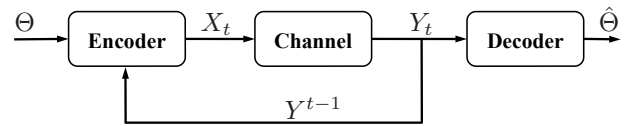}
\caption{A noisy memoryless channel with a noiseless causal feedback link.}
\label{fig:feedback}
\end{figure}

Consider the problem of coding over a discrete memoryless channel (DMC) 
with noiseless feedback as depicted in Fig.~\ref{fig:feedback}.
The DMC is described by finite input and output sets~$\mathcal{X}$ and~$\mathcal{Y}$, and
a collection of conditional probabilities $P(Y|X)$. To simplify notation, and without loss of generality, we assume that 
\begin{align}
\mathcal{X} &= \{0,1,\ldots, |\mathcal{X}|-1\}, \\
\mathcal{Y} &= \{0,1,\ldots, |\mathcal{Y}|-1\}.
\end{align}
%

Let $C$ denote the \emph{Shannon capacity} of the DMC $P(Y|X)$ \cite[p.~184]{CoverBook}:
\begin{align}\label{eq:C}
C=\max_{P_X} I(X;Y),
\end{align}
and let $(\pi^\star_0, \pi^\star_1, \ldots, \pi^\star_{|\mathcal{X}|-1})$ be the maximizer of \eqref{eq:C}, the so-called \emph{capacity-achieving input distribution}.  The operational meaning of the Shannon capacity is discussed in Section~\ref{sec:rate-reliability}.

The following result will be used in our proofs. 
\begin{fact}[Theorem 4.5.1 in \cite{Gallager68}] 
\label{Gallager} 
Consider a DMC with capacity-achieving input distribution $\pi^\star_0, \pi^\star_1, \ldots, \pi^\star_{|\mathcal{X}|-1}$.
For each $k\in\{0,1,\ldots, |\mathcal{X}|-1\}$, if $\pi^\star_k > 0$,
\begin{align*}
D\bigg(P(Y|X=k) \bigg\| \sum \limits_{l=0}^{|\mathcal{X}|-1} \pi^\star_l P(Y|X=l)\bigg) = C.
\end{align*}
\end{fact}

Let $C_1$ be the KL~divergence between the two most distinguishable inputs of the DMC: 
\begin{align}
C_1= \max \limits_{x,x' \in \mathcal{X}} D(P(Y|X=x) \| P(Y|X=x')).
\end{align} 
We also denote
\begin{align}\label{eq:C2}
C_2=\max_{y \in \mathcal{Y}} \frac{\max_{x \in \mathcal{X}} P(Y=y|X=x)}{\min_{x \in \mathcal{X}} P(Y=y|X=x)}.
\end{align} 
%
In this paper, we assume $C$, $C_1$, $C_2$ are positive and finite.\footnote{It can be easily shown that 
$C \le C_1 \le \log C_2 \le C_2$. Furthermore, if $C_1<\infty$, then the transition probability $P(Y=y|X=x)$ is positive for all $x\in\mathcal{X}$ and $y\in\mathcal{Y}$, which implies that $C_2<\infty$ as well.
Therefore, $C>0$ and $C_1<\infty$ are sufficient to ensure that 
$C$, $C_1$, $C_2$ are positive and finite.
}

Let $\tau$ denote the total transmission time (or equivalently the total length of the code). 
The transmitter wishes to communicate a message $\Theta$ to the receiver, where the message is uniformly distributed over a message set
 \begin{equation}\label{eq:messageset}
 \Omega := \{1,2,\ldots,M\}.
 \end{equation} 
 To this end, the transmitter produces channel inputs $X_t$ for $t=0,1,\ldots,\tau-1$, 
 which it can compute as a function of the message $\Theta$ and (thanks to the noiseless feedback) 
 also of the past channel outputs $Y^{t-1}:=(Y_0,Y_1,\ldots,Y_{t-1})$: 
\begin{equation}\label{eq:13}
X_t= e_t(\Theta, Y^{t-1}), \quad t=0,1,\ldots,\tau-1,
\end{equation}
for some encoding function $e_t\colon \Omega \times \mathcal{Y}^{t} \to \mathcal{X}$.

After observing the $\tau$ channel outputs $Y_0,Y_1,\ldots,Y_{\tau-1}$, the receiver guesses the message $\Theta$ as  
\begin{equation}\label{eq:2}
\hat{\Theta} = d\big( Y^{\tau-1}\big),
\end{equation} 
for some decoding function $d\colon \mathcal{Y}^{\tau} \to \Omega$. 
The probability of error of the scheme is thus
\begin{equation*}
\Pe := \textnormal{Pr}(\hat{\Theta} \neq \Theta).
\end{equation*}

In contrast to fixed-length coding where 
the total transmission time $\tau$ is deterministic and known before the transmission starts,
in this paper, our focus is on variable-length coding, 
i.e., the case where $\tau$ is a random stopping time decided at the receiver 
as a function of the observed channel outputs. 
Thanks to the noiseless feedback, the transmitter is also informed of 
the channel outputs and hence of the stopping time.

For a fixed DMC and for a given $\epsilon>0$, the goal is to find encoding and decoding rules as in \eqref{eq:13} and \eqref{eq:2}, and a stopping time $\tau_\epsilon$ such that the probability of error satisfies $\Pe\le \epsilon$ and the expected number of channel uses $\mathbb{E}[\tau_\epsilon]$ is minimized.
Let $\mathbb{E}[\tause]$ be the minimum expected number of channel uses that can be achieved by coding schemes with the stopping rule $\tau_{\epsilon}$.

We shall often use the functions $ \{\gamma_{y^{t-1}}\}$ for $y^{t-1}\in\mathcal{Y}^{t}$ and $t\in\{0,1,\ldots, \tau-1\}$  where
\begin{subequations}\label{eq:gamma}
\begin{IEEEeqnarray}{rCl}
\gamma_{y^{t-1}}  \colon  \Omega & \to & \mathcal{X}\\
i & \mapsto & e_t(i, y^{t-1})
\end{IEEEeqnarray}
\end{subequations}
to describe the encoding process.
To simplify notation and where it is clear from the context, we shall often omit the subscript $y^{t-1}$ and simply write $\gamma$.

In some examples we also allow for \emph{randomized} encoding rules. In this case the encoding is described by the \emph{random} encoding functions $\{\Gamma_{y^{t-1}}\}$ whose realizations $\gamma_{y^{t-1}}$ are of the form in~\eqref{eq:gamma}. Again, for notational convenience we shall omit the subscript $y^{t-1}$ where it is clear from the context.

Note that a variable-length code differs from a single encoding function; rather, it is an adaptive rule that dictates the choice of (random) encoding functions depending on the past channel observations and 
past selected encoding functions prior to the stopping time. 
In this paper, we refer to  this adaptive rule as an encoding scheme, $\fc$, 
which together with the particular realization of channel outputs $y_0,y_1,\ldots, y_{\tau_{}-2}$, dictates the encoding functions 
$\Gamma^{\fc}_{y^0}, \Gamma^\fc_{y^1}, \ldots, \Gamma^\fc_{y^{\tau_{}-2}}$.

\subsection{Asymptotic Bounds on Minimum Expected Length}
In \cite{Burnashev76}, Burnashev provided the following lower and upper bounds on the 
minimum expected number of channel uses, $\mathbb{E}[\tause]$, for a large class of DMCs and arbitrary $\epsilon>0$.

\begin{fact}[Theorems 1 and 2 in \cite{Burnashev76}]
\label{Fact:Burnashev}
For any DMC with $C>0$ and $C_1 < \infty$: 
\begin{align}\label{FactLB}
\mathbb{E}[\tause] &\ge \left(\frac{\log M}{C} + \frac{\log\frac{1}{\epsilon}}{C_1}\right)(1-o(1)),
\end{align}
and
\begin{align}
\label{FactUB}
\mathbb{E}[\tause] &\le 
\left( \frac{\log M}{C} + \frac{\log\frac{1}{\epsilon}}{C_1} \right) (1+o(1))
\end{align}
where $o(1)\to 0$ as $\epsilon \to 0$.\footnote{If $\epsilon \to 0$, then $o(1)\to 0$ regardless of $M$ being fixed or $M \to \infty$. For fixed $\epsilon$, $\mathbb{E}[\tause] \approx \frac{(1-\epsilon)\log M}{C}$ and hence, the positive term $o(1) \not\to 0$ even if $M \to \infty$ (see \cite{Polyanskiy11} for more details).} 
\end{fact}

Inequality~\eqref{FactLB} was proved in \cite{Burnashev76} using a Martingale argument,
and it was reproved  more concisely in \cite{Berlin09}. A strictly tighter 
version of~\eqref{FactLB} was provided in \cite{Polyanskiy11}. 

Burnashev proved the upper bound~\eqref{FactUB} using the following two-phase scheme  \cite{Burnashev76}.
While in the first phase (\emph{communication phase}) the transmitter iteratively refines the receiver's belief about the true message, in the second phase (\emph{confirmation phase}) it simply 
confirms whether the receiver's highest belief after the first phase corresponds to the true message. 
{As} shown in \cite{Yamamoto79, Caire06}  the specific scheme 
in the first phase can be exchanged by any capacity achieving block coding schemes.

\subsection{Stochastic Control View}
\label{StochView}

%
%
\begin{figure}[htp]
\centering
\psfrag{g}{$\gamma$}
\psfrag{f}{\hspace*{-.33in}{\small{$X_t=\gamma(\Theta)$}}}
\psfrag{t}{{$\Theta$}}
\psfrag{x}{{$X_t$}}
\psfrag{y}{{$Y_t$}}
\psfrag{z}{{$Y^{t-1}$}}
\psfrag{h}{{$\hat{\Theta}$}}
\includegraphics[width=0.489\textwidth]{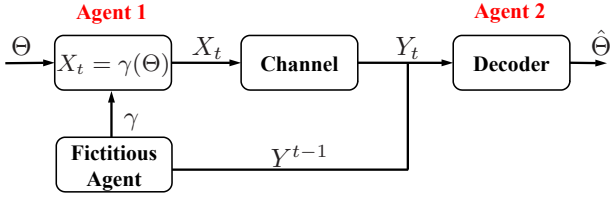}

\caption{Two-agent problem with common and private observations from the point of view of the fictitious agent. }
\label{fig:controller}
\end{figure}

The problem of variable-length coding with noiseless feedback is
a decentralized team problem with two agents (the encoder and the decoder) and non-classical information structure {\cite{Witsenhausen1968}}. {Appealing} to \cite{Mahajan08}, the problem
can be interpreted as a special case of
active hypothesis testing  \cite{HypJournal} in which a (fictitious) Bayesian decision maker is responsible to enhance his information 
about the correct message in a speedy manner  
by sequentially sampling from conditionally independent observations 
at the output of the channel {(given the input)}. 
Here the decision maker 
has access to the channel output symbols causally (common observations) 
and is responsible to control the conditional distribution of the observations given the 
true message (private observation) by selecting encoding functions for the encoder
which map the message $\Theta$ to the input symbols of the channel. In other words, as also observed in \cite{toddISIT},
the problem can be viewed as a (centralized) partially observable Markov decision problem (POMDP) 
with (static) state space $\Omega$ and
the observation space $\mathcal{Y}$. Let $\mathcal{E}:=~\left\{\gamma(\cdot):\Omega \to \mathcal{X} \right\}$ be the set of all mappings from $\Omega$ to $\mathcal{X}$. 
The action space (for the fictitious agent) becomes $\mathcal{E} \cup \{{T}\}$ 
where ${T}$ denotes the termination of the transmission phase, hence the realization of the stopping time~$\tau$.  

Casting the problem as a POMDP allows for the structural characterization of the information state, also known as sufficient statistics:
{ Let t}he decision maker's belief about each possible message $i\in\Omega$, updated
after each channel use (observation) for $t=0,1,\ldots,\tau-1$, be 
\begin{equation}
\rho_i(t):= \textnormal{Pr}(\Theta=i| Y^{t-1}).
\end{equation}
{The decision maker's posteriors about the messages collectively,} 
\begin{equation}
\boldsymbol{\rho}(t):= [\rho_1(t), \rho_2(t), \ldots, \rho_M(t)],
\end{equation}
form a sufficient statistics for our Bayesian decision maker. Furthermore, this decision maker's posterior at any time $t$ coincides with the receiver's posterior and, thanks to the perfect feedback, is available to the transmitter. 
(Notice that $\rho_i(0)=\textnormal{Pr}(\Theta=i)=\frac{1}{M}$ denotes the receiver's initial belief of $\Theta=i$ before the transmission starts.)  In other words, the selection of encoding and decoding rules as a function
of this posterior does not incur any loss of optimality \cite{Bertsekas}. In particular, the optimal 
receiver produces as its guess the message with the highest posterior at time $\tau$, i.e.,
\begin{align}
\hat{\Theta} = \argmax_{i\in \Omega} \rho_i(\tau). \label{ML_decoder}
\end{align}
{We also note that the dynamics of the information state, i.e., the posterior, follows Bayes' rule. More specifically, given 
an encoding function $\gamma$ at time $t$ and an information state $\boldsymbol{\rho}$, the conditional distribution of the 
next channel output $Y_t$, given the past observation $Y^{t-1}$, is 
$$P_{\boldsymbol{\rho}}(y)=\sum_{i=1}^M \rho_i P(Y=y|X=\gamma(i)).$$  Similarly,  given also the output symbol 
$Y_t=y$, according to Bayes' rule,  the posterior at time $t+1$ is:
\begin{align*}
\boldsymbol{\rho}(t+1) = \Big[\frac{\rho_1 P_{\gamma(1)}(y)}{P_{\boldsymbol{\rho}}(y)},\ldots,\frac{\rho_M P_{\gamma(M)}(y)}{P_{\boldsymbol{\rho}}(y)}\Big].
\end{align*}}

Taking cue from the
seminal work of DeGroot on statistical decision theory \cite{DeGroot70}, 
the above stochastic control view of the variable-length coding has been used in \cite{ISIT2012} 
to characterize the performance of any given coding scheme using the information utility 
provided by the channel output.  {Information utility, here, 
generalizes the Shannon theoretic notion of mutual information\cite{DeGroot70},  \cite{ISIT2012}. 
More specifically, consider any given measure
of the uncertainty of the posterior vector; information utility is defined as 
the expected reduction in the uncertainty of the posterior at time $t+1$ relative to that at time $t$.}
{The result in \cite{ISIT2012}, as also manifested in Lemma~\ref{lem:propEJS},  
implies a characterization of the performance of a given coding scheme in terms of the symmetric divergences JS and EJS 
between the conditional  output distributions of the channel induced by the encoding function}. 
In particular, taking the average log-likelihood as a measure of uncertainty, under any encoding function $\gamma\colon \Omega \to \mathcal{X}$ used
at time $t$ over a DMC $P(Y|X)$, one can quantify the expected reduction in uncertainty in form of 
\begin{align}
\label{EJSgammadet}
EJS(\boldsymbol{\rho}(t), \gamma)
&:= EJS\big(\boldsymbol{\rho}(t);P_{\gamma(1)},\ldots,P_{\gamma(M)}\big).
\end{align}

In the sections that follow, 
we utilize this connection, non-negativity of EJS, and a submartingale level crossing theorem as the basis of our achievability analysis. 
In particular, in Section~\ref{Sec:Main} we specificize the 
approach in \cite{ISIT2012} with respect to the EJS divergence induced by the encoding mapping. This allows us to 
provide achievability analysis for two one-phase coding schemes, namely variable-length posterior matching and MaxEJS. 
{These schemes are based on the suboptimal stopping rule described in the next section.}
Furthermore, we show that MaxEJS coding scheme provably achieves Burnashev's asymptotic optimal 
performance given by \eqref{FactUB}. 

\subsection{{A Suboptimal Stopping Rule}}\label{sec:decoding}

In this paper we focus on the following (possibly suboptimal) stopping rule.
For any given coding scheme $\fc$, 
the transmission is only stopped when  one of the posteriors becomes larger than $1-\epsilon$, where $\epsilon>0$ is the desired probability of error:
\begin{align}\label{eq:deftau}
\taute:=\min \{t: \max_{i \in \Omega} \rho_i(t)\ge 1-\epsilon\}. 
\end{align}

From the described optimal decoding rule of \eqref{ML_decoder}, the constraint on the probability of error is satisfied by any coding scheme with the stopping rule \eqref{eq:deftau}:
\begin{align*}
\Pe = \mathbb{E} [1-\max_{i \in \Omega} \rho_i(\tilde\tau_\epsilon)]\le \epsilon.
\end{align*}

\section{Main Result and Applications}
\label{Sec:Main}

In this section, we first characterize the performance of an encoding scheme in terms of 
its corresponding extrinsic Jensen--Shannon (EJS) divergence obtained.  To make this  precise we first introduce 
some further notation to allow for randomized encoding. 

\ignore{
Given a DMC $P(Y|X)$ and a (deterministic) encoding function $\gamma\colon \Omega \to \mathcal{X}$ together with a set of time-$t$ posteriors $\boldsymbol{\rho}(t)$, we use the shorthand notation:
\begin{align}
EJS(\boldsymbol{\rho}(t), \gamma)
&:= EJS\big(\boldsymbol{\rho}(t);P_{\gamma(1)},\ldots,P_{\gamma(M)}\big).
\end{align} }

For a (possibly) randomized encoding rule $\Gamma$, we use the shorthand notation:
\begin{align}
\label{EJSgamma}
EJS(\boldsymbol{\rho}(t), \Gamma)
&:= \sum_{\gamma\in\mathcal{E}} \Pr (\Gamma=\gamma|Y^{t-1}) EJS (\boldsymbol{\rho}(t), \gamma)
\end{align}
where recall that $\mathcal{E}$ denotes the set of all possible encoding functions, and $EJS (\boldsymbol{\rho}(t), \gamma)$ is defined in \eqref{EJSgammadet}. 


\subsection{{Main Theorem}}

Let 
\begin{equation}
\tilde{\rho}:=1-\frac{1}{1+\max\{\log M,\log\frac{1}{\epsilon}\}}.
\end{equation}
\begin{theorem}
\label{Thm:EJS-LB}
Consider a (possibly randomized) encoding scheme $\fc$ under which
at each time $t=0,1,\ldots, \tilde{\tau}_{\epsilon}-1$ and for each $y^{t-1}$ the encoding function 
$\Gamma^\fc$ satisfies
\begin{subequations} \label{CondTestThm}
\begin{align}
EJS(\boldsymbol{\rho}(t),\Gamma^\fc) \geq R_\tmin,
\end{align}
and furthermore, 
\begin{align}
EJS(\boldsymbol{\rho}(t),\Gamma^\fc) \geq \tilde{\rho} E_\tmin \qquad \text{if } \max_{i\in\Omega} \rho_i(t) \geq \tilde{\rho},
\end{align}
\end{subequations}
for  some $E_\tmin \ge R_\tmin > 0$.
Then, 
\begin{IEEEeqnarray}{rCl}
\mathbb{E}_\fc[\taute]\le \frac{\log M + \log\log\frac{M}{\epsilon} }{R_\tmin} +\frac{\log\frac{1}{\epsilon}+1}{E_\tmin} + \frac{6(4C_2)^2}{R_\tmin E_\tmin}\IEEEeqnarraynumspace
\end{IEEEeqnarray}
where $C_2$ is defined in~\eqref{eq:C2}.
\end{theorem}
\begin{corollary}\label{cor1}
Under the assumptions of Theorem~\ref{Thm:EJS-LB},
\begin{IEEEeqnarray}{rCl}
\mathbb{E}_\fc[\taute]\le \left(\frac{\log M}{R_\tmin} +\frac{\log\frac{1}{\epsilon}}{E_\tmin}\right)(1+o(1))
\end{IEEEeqnarray}
where $o(1)\to 0$ as $\epsilon \to 0$ or $M \to \infty$.
\end{corollary}
The proof of Theorem~\ref{Thm:EJS-LB} is given in Appendix~\ref{App:Thm}
and is based on  the following 
fact about submartingales: For any submartingale $\{\xi(t)\}$ with respect to a filtration $\{\mathcal{F}(t)\}$, $t=0,1,2,\ldots$, if 
there exist positive constants $K_1$ and $K_2$ such that
\begin{subequations} \label{eqn12}
\begin{align}
& \mathbb{E} [ \xi(t+1) | \mathcal{F}(t) ] \ge \xi(t) +K_1 \hspace{0.1 in} {\mbox{if}} \hspace{0.1 in} \xi(t) < 0, \label{eqn1} \\   
& \mathbb{E} [ \xi(t+1) | \mathcal{F}(t) ] \ge \xi(t) +K_2 \hspace{0.1 in} {\mbox{if}} \hspace{0.1 in} \xi(t) \ge 0, \label{eqn2}
\end{align}
\end{subequations}
then, under certain 
technical conditions, the stopping time $\upsilon = \min \{ t: \xi(t) \ge B \}$, $B>0$ can be approximately upper bounded as 
$$\mathbb{E} [\upsilon] \lesssim  \frac{B - \xi(0)}{K_2} + \xi(0) {\bf{1}}_{\{\xi(0) <0\}} \left(\frac{1}{K_2} - \frac{1}{K_1} \right).$$

Now let $\mathcal{F}(t)$ denote the history of the receiver's knowledge up to time $t$, i.e., 
$\mathcal{F}(t)=\sigma \{ Y^{t-1} \}$, and let
\begin{align*}
\tilde{U}(t) :=& - U(t) - \log\frac{\tilde{\rho}}{1-\tilde{\rho}}\\
=& \sum_{i=1}^M \rho_i(t) \log \frac{\rho_i(t)}{1-\rho_i(t)} - \log\frac{\tilde{\rho}}{1-\tilde{\rho}}.
\end{align*}
From Lemma~2, 
\begin{align}
\mathbb{E}_\fc \left [\tilde{U}(t+1) | \mathcal{F}(t) \right ] = \tilde{U}(t) + EJS(\boldsymbol{\rho}(t),\Gamma^\fc),
\end{align}
and hence the sequence $\{\tilde{U}(t)\}$ forms a submartingale. The assertion of the theorem directly follows from~\eqref{eqn12} when setting $K_1= R_{\min}$ and $K_2= \tilde{\rho} E_{\min}$.

\subsection{Application I: Variable-Length Posterior Matching}
\label{subsec:pm}

We consider a variable-length version of the coding schemes in \cite{Horstein63, Burnashev74, Shayevitz11}. At each time $t=0,1,\ldots, \taute-1$, if $\Theta=i$ and given the posterior vector $\boldsymbol{\rho}(t)$, 
 the input $X(t)$ takes value in the set
\begin{align*}
\lefteqn{\mathcal{X}_{i}(t)
:=\bigg\{ x\in \mathcal{X} \colon \sum_{i'=1}^{i-1} \rho_{i' }(t) <  \sum_{x' \leq x} \pi_{x'}^\star} \\
&\hspace*{1.83in} \text{ and } \sum_{x' < x} \pi_{x'}^\star \leq  \sum_{i'=1}^{i} \rho_{i' }(t)\bigg\};
\end{align*}
where each value $x\in\mathcal{X}_i(t)$ is taken with probability
\begin{align}
\label{ProbPMXt}
\lefteqn{\Pr \big ( X(t) = x| \Theta=i, Y^{t-1}=y^{t-1}\big )}\nonumber \\ 
 &= \hspace*{-.026in} \frac{\min \Big\{ \sum\limits_{i'=1}^{i} \rho_{i' }(t), \sum\limits_{x' \leq x} \pi_{x'}^\star \Big\} - \max  \Big\{ \sum\limits_{i'=1}^{i-1} \rho_{i' }(t),  \sum\limits_{x' < x} \pi_{x'}^\star \Big\}}{\rho_i(t)}.
\end{align}

\begin{figure*}[htp!]
\centering
\psfrag{a}{$\pi^{\star}_0$}
\psfrag{b}{\hspace*{-0.05in}$\pi^\star_1$}
\psfrag{c}{\hspace*{-0.1in}$\pi^\star_{k-1}$}
\psfrag{d}{\hspace*{-0.05in}$\pi^\star_k$}
\psfrag{e}{\hspace*{-0.1in}$\pi^\star_{k+1}$}
\psfrag{f}{{\hspace*{-0.1in}$\pi^\star_{|\mathcal{X}|-1}$}}
\psfrag{g}{{\hspace*{-0.11in}$\rho_1(t)$}}
\psfrag{h}{{\hspace*{-0.25in}$\rho_2(t)$}}
\psfrag{i}{{\hspace*{-0.15in}$\rho_{i-1}(t)$}}
\psfrag{j}{{\hspace*{-0.15in}$\rho_i(t)$}}
\psfrag{k}{{\hspace*{-0.03in}$\rho_{i+1}(t)$}}
\psfrag{m}{{\hspace*{-0.05in}$\rho_M(t)$}}
\psfrag{n}{{\hspace*{-0.23in}$\hat{\rho}_{i,k-1}(t)$}}
\psfrag{o}{{\hspace*{-0.1in}$\hat{\rho}_{i,k}(t)$}}
\psfrag{p}{{\hspace*{-0.15in}$\hat{\rho}_{i,k+1}(t)$}}
\includegraphics[width=1\textwidth]{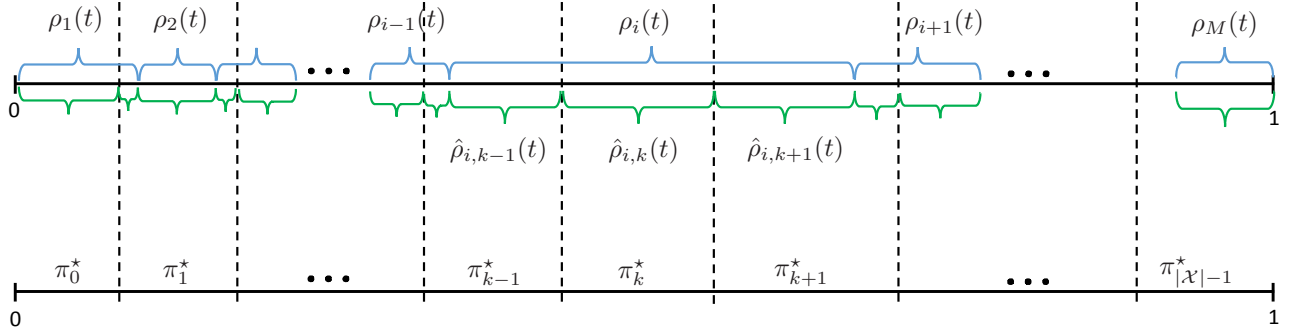}
\caption{Posterior Matching scheme for a DMC with capacity-achieving input distribution $\pi^\star_0, \pi^\star_1, \ldots, \pi^\star_{|\mathcal{X}|-1}$. In this example, $\mathcal{X}_i(t)=\{k-1,k,k+1\}$ since $\sum_{i'=1}^{i-1} \rho_{i' }(t) <  \sum_{x' \leq x} \pi_{x'}^\star$ for all $x \ge k-1$ and $\sum_{x' < x} \pi_{x'}^\star \leq  \sum_{i'=1}^{i} \rho_{i' }(t)$ for all $x\le k+1$. It is clear that as $\rho_i(t)$ approaches 1, the candidate set $\mathcal{X}_i(t)$ gets larger and given that $\Theta=i$, the posterior matching scheme selects the channel input $x$ out of this set with probability $\hat{\rho}_{i,x}(t) / \rho_i(t)$ which converges to $\pi^\star_x$.}
\label{fig:PMsets}
\end{figure*}

Let $\hat{\rho}_{i,x}(t)$ denote the numerator in the right-hand side of \eqref{ProbPMXt}. Fig.~\ref{fig:PMsets} shows an example on how posterior matching scheme selects channel inputs.

\begin{proposition}\label{prop:PM}
Under the above variable-length posterior matching encoding\footnote{Assumption $C_1 < \infty$ circumvents the fixed point phenomena under which the posterior matching scheme cannot achieve any positive rate.}, and for each $t =0,1,\ldots, \taute-1$
and all possible output sequences $y^{t-1}$, 
$$EJS(\boldsymbol{\rho}(t), \Gamma^{\mathrm{PM}}) \geq  C.$$
\end{proposition}
The proof of Proposition~\ref{prop:PM} is given in Appendix~\ref{app:PM}.

\allowdisplaybreaks[4]

Proposition~\ref{prop:PM} implies that the variable-length posterior matching encoding satisfies \eqref{CondTestThm} with $R_\tmin= E_\tmin= C$.
\begin{remarks}
By Theorem~\ref{Thm:EJS-LB} and Proposition~\ref{prop:PM}, 
under the variable-length posterior matching encoding 
\begin{equation}
\mathbb{E}_{\Gamma^\mathrm{PM}}[\taute]\le \frac{\log M+\log\frac{1}{\epsilon}+1+\log\log\frac{M}{\epsilon} }{C}+ \frac{6(4C_2)^2}{C^2}.
\end{equation}
\end{remarks}

\subsection{Application II: MaxEJS Coding}
\label{Sec:HeuEJS}

We present a new coding scheme based on the greedy maximization of EJS divergence.
At each time $t=0,1,\ldots, \taute-1$ and given the posterior vector $\boldsymbol{\rho}(t)$, MaxEJS chooses the $\gamma^*$ that maximizes the EJS divergence: 
\begin{equation}\label{eq:gammamax}
\gamma^*:=\argmax_{\gamma\in\mathcal{E}}  EJS(\boldsymbol{\rho}(t),\gamma).
\end{equation}

\begin{proposition}
\label{Prop:EJSmax}
For every $t =0,1,\ldots, \taute-1$ and all possible output sequences $y^{t-1}$,
MaxEJS encoding satisfies 
\begin{subequations}
\begin{align}
EJS(\boldsymbol{\rho}(t),\gamma^*) \geq C, \label{eqn1C}
\end{align}
and furthermore,
\begin{align}
EJS(\boldsymbol{\rho}(t),\gamma^*) \geq \tilde{\rho} C_1 \qquad \text{if } \max_{i\in\Omega} \rho_i(t) \ge \tilde{\rho}. \label{eqn2C1}
\end{align}
\end{subequations}
\end{proposition}
The proof of Proposition~\ref{Prop:EJSmax} is given in Appendix~\ref{app:EJSMAX}.
\begin{remarks}
By Theorem~\ref{Thm:EJS-LB} and Proposition~\ref{Prop:EJSmax}, 
\begin{IEEEeqnarray}{rCl}
\mathbb{E}_{\Gamma^{\textnormal{MaxEJS}}}[\taute]\le  \frac{\log M + \log\log\frac{M}{\epsilon} }{C} +\frac{\log\frac{1}{\epsilon}+1}{C_1} + \frac{6(4C_2)^2}{C C_1},\nonumber \\
\end{IEEEeqnarray}
and thus
MaxEJS encoding together with the decoding and stopping rules described in \eqref{ML_decoder} and \eqref{eq:deftau} 
achieves Burnashev's optimal asymptotic performance in~\eqref{FactUB}, see Corollary~\ref{cor1}.
\end{remarks}

\begin{remarks}
The presented deterministic one-phase scheme differs from the previous schemes 
achieving Burnashev's optimal asymptotic performance, which are randomized and have two phases
\cite{Burnashev76,Yamamoto79,Ooi98,Caire06}. However, (\ref{eqn1C}) and (\ref{eqn2C1}) show that 
this one-phase scheme operationally moves between the two regimes of \emph{communication} and \emph{confirmation}.
\end{remarks}

The computational complexity of the MaxEJS coding scheme could be prohibitive. In Section~\ref{sec:BinaryOpt},
 we propose simpler coding schemes for a class of binary-input channels that achieve Burnashev's optimal asymptotic performance in~\eqref{FactUB}.

\section{Coding for {Symmetric Binary-Input} Channels}
\label{Sec:Sym}

In this subsection, we focus on channels with binary inputs $\mathcal{X}=\{0,1\}$ and with the following property
\begin{equation}\label{eq:channelcond}
P(Y=y|X=0)=P(Y=f(y)|X=1), \quad \forall y\in\mathcal{Y}\end{equation}
for a permutation $f: \mathcal{Y} \to \mathcal{Y}$ where $f = f^{-1}$, i.e., $f$ is its own inverse.


The first attempt to address the problem of coding over a symmetric binary-input channel goes back to Horstein's coding 
scheme~\cite{Horstein63} over a binary symmetric channel (BSC) with a crossover probability $p \in (0,1/2)$. 
Horstein considered the message to be a point in the interval $[0,1]$ and suggested that to 
achieve the capacity of the channel, at any given time the transmitter 
selects the input of the channel such as to signal to the receiver whether the 
message is smaller than the median of the posterior or larger.  Later, Burnashev and Zigangirov \cite{Burnashev74}, presented a similar (randomized) coding scheme for discrete message sets as in~\eqref{eq:messageset} and proved that this scheme achieves capacity.

 In Section~\ref{Sec:BSC}, we present and analyze a \emph{deterministic} scheme for arbitrary symmetric binary-input channels satisfying~\eqref{eq:channelcond}, {which resembles the Burnashev-Zigangirov scheme, when specialized to the BSC}. In Section~\ref{sec:BinaryOpt}, we then improve our scheme so that it 
achieves Burnashev's optimal asymptotic performance in~\eqref{FactUB} over this class of symmetric binary-input channels.


\subsection{{Generalized} Horstein-Burnashev-Zigangirov Scheme}
\label{Sec:BSC}

{Our generalization of the Horstein-Burnashev-Zigangirov scheme is deterministic.}
For each time {$t=0,1,\ldots, \taute-1$} and given the posterior vector $\boldsymbol{\rho}(t)$, we choose the encoding function:
\begin{equation}
\gamma^{\textnormal{GHBZ}}(i) = \begin{cases} 0 & 1\leq i\leq k^*\\1& k^* < i \leq M \end{cases}
\end{equation} 
where
\begin{equation}
k^*:=\argmin_{k\in\Omega} \Big|\sum_{i=1}^{k} \rho_i(t) - \frac{1}{2}\Big|.
\end{equation}


\begin{proposition}
\label{Prop:DscrtH}
Consider the deterministic scheme proposed above over a binary-input DMC that satisfies~\eqref{eq:channelcond}. For every $t=0,1,\ldots, \taute-1$ 
and all possible output sequences $y^{t-1}$,
\begin{equation}\label{eq:maxC}
EJS(\boldsymbol{\rho}(t), \gamma^{\textnormal{GHBZ}}) \geq  C.
\end{equation}
\end{proposition}
The proof is given in Appendix~\ref{app:proofHorstein}. 
\begin{remarks}
By Theorem~\ref{Thm:EJS-LB} and Proposition~\ref{Prop:DscrtH}, the described encoding satisfies
\begin{equation}
\label{binC}
\mathbb{E}_{\gamma^{\textnormal{GHBZ}}}[\taute]\le  \frac{\log M+\log\frac{1}{\epsilon}+1+\log\log\frac{M}{\epsilon} }{C}+ \frac{6(4C_2)^2}{C^2}.
\end{equation}
\end{remarks}

Notice that, when specialized to a binary-input channel, the variable-length posterior matching scheme of Section~\ref{subsec:pm}, at each time $t=0,1,\ldots, \taute-1$ and {given the posterior vector $\boldsymbol{\rho}(t)$}, 
 chooses encoding function $\gamma^{\textnormal{GHBZ}}$ with probability 
\begin{align} \label{gamma}
\lambda_{\gamma^{\textnormal{GHBZ}}}= \frac{\delta_2(t)}{\delta_1(t)+\delta_2(t)}
\end{align}
where 
\begin{align}
\label{delta12}
\delta_1(t):=\bigg|\sum_{i=1}^{k^*} \rho_i(t)-\frac{1}{2}\bigg|, \quad \delta_2(t):=\bigg|\sum_{i=1}^{k^*_2} \rho_i(t)-\frac{1}{2}\bigg|,
\end{align}
and
\begin{align}
\label{k*2}
k^*_2:=k^*- \text{sign} \bigg (\sum_{i=1}^{k^*} \rho_i(t)-\frac{1}{2} \bigg );
\end{align}
and it chooses the encoding function 
\begin{equation}
\bar{\gamma}^{\textnormal{GHBZ}}(i) = \begin{cases} 0 & 1\leq i\leq k_2^*\\1& k_2^* < i \leq M \end{cases}
\end{equation}
with probability $\bar{\lambda}_{\gamma^{\textnormal{GHBZ}}}=1- \lambda_{\gamma^{\textnormal{GHBZ}}}$. 

Combining Proposition~\ref{Prop:DscrtH} with Proposition~\ref{prop:PM}, we have that there exists a class (a continuum) of randomized schemes that satisfy~\eqref{eq:maxC}:
\begin{corollary}
\label{generalize-horstein}
Every (randomized) encoding function $\Gamma$ that selects $\gamma^{\textnormal{GHBZ}}$ with probability $\lambda\ge\lambda_{\gamma^{\textnormal{GHBZ}}}$ in (\ref{gamma}) and selects $\bar{\gamma}^{\textnormal{GHBZ}}$  with 
probability $\bar{\lambda}=1-\lambda$, satisfies \eqref{CondTestThm} with $R_\tmin= E_\tmin= C$. 
\end{corollary}
This corollary provides an alternative proof that Burnashev and Zigangirov's variable-length coding scheme  \cite{Burnashev74} satisfies~\eqref{binC} over the BSC with crossover probability $p\in(0,1/2)$.
In fact, their scheme selects $\gamma^{\textnormal{GHBZ}}$ and $\bar{\gamma}^{\textnormal{GHBZ}}$ with probabilities
$\lambda=\frac{\nu(\delta_2(t))}{\nu(\delta_1(t))+\nu(\delta_2(t))}$ and $\bar{\lambda}=1-\lambda$, respectively, where $\nu(x)= \log\frac{0.5+(1-2p)x}{0.5-(1-2p)x}$.
{
We next prove that $\frac{\nu(\delta_2(t))}{\nu(\delta_1(t))+\nu(\delta_2(t))} \ge \frac{\delta_2(t)}{\delta_1(t)+\delta_2(t)}$, which by Corollary~\ref{generalize-horstein} establishes that the Burnashev-Zigangirov scheme indeed satisfies~\eqref{binC}. 

Notice that $\nu(x)=\log\left( -1 + \frac{1}{0.5-(1-2p)x}\right)$ is convex for all $x$ because $p\in(0,1/2)$. Since  also $f\colon x \mapsto \frac{\nu(x)}{\nu(\delta_2(t))}$ is convex and since $f(0) =0$ and $f(\delta_2(t))=1$, we conclude that $\frac{\nu(x)}{\nu(\delta_2(t))} \le \frac{x}{\delta_2(t)}$, for all $x\in[0, \delta_2(t)]$. By~\eqref{delta12} and \eqref{k*2}, $0\leq \delta_1(t)\leq \delta_2(t)$ and hence
$\frac{\nu(\delta_1(t))}{\nu(\delta_2(t))} \le \frac{\delta_1(t)}{\delta_2(t)}$. This immediately establishes the desired inequality~$\frac{\nu(\delta_2(t))}{\nu(\delta_1(t))+\nu(\delta_2(t))} \ge \frac{\delta_2(t)}{\delta_1(t)+\delta_2(t)}$. }

\subsection{Optimal Binary Variable-Length Codes}
\label{sec:BinaryOpt}

Motivated by the analysis above, we strive to simplify our deterministic one-phase  MaxEJS scheme for the simpler symmetric binary-input channels.  We propose the following encoding scheme. At each time $t=0,1,\ldots, \taute-1$ and each sequence of observations $Y^{t-1}=y^{t-1}$, we choose the encoding function $\gamma$ in a way that for all $i \in \{j\in\Omega \colon \gamma(j)=0\}$, 
\begin{equation}\label{eq:detbinarycond}
0 \leq \sum_{j\in\Omega \colon \gamma(j)=0} \rho_j(t) - \sum_{j\in\Omega \colon \gamma(j)=1} \rho_j(t) < \rho_i(t).
\end{equation}


By condition \eqref{eq:detbinarycond}, at each time $t$, the probabilities of sending a 0 or a 1 are approximately $(1/2, 1/2)$ when all posteriors $\{\rho_i(t)\}_{i\in\Omega}$ are small, and they are 
$(\max_{i\in\Omega} \rho_i(t), 1 - \max_{i\in\Omega} \rho_i(t))$ 
when $\max_{i\in\Omega} \rho_i(t)$ is larger than $1/2$.

\begin{proposition}
\label{Prop:BSC}
If for every $t=0,1,\ldots, \taute-1$ and every sequence of observations $Y^{t-1}=y^{t-1}$ the encoding function $\gamma$ satisfies~\eqref{eq:detbinarycond}, then
\begin{subequations}
\begin{align}
EJS(\boldsymbol{\rho}(t),\gamma) \geq C,
\end{align}
and 
\begin{align}
EJS(\boldsymbol{\rho}(t),\gamma) \geq \tilde{\rho} C_1 \qquad \text{if } \max_{i\in\Omega} \rho_i(t) \ge \tilde{\rho}.
\end{align}
\end{subequations}
\end{proposition}

The proof is given in Appendix~\ref{app:proofbinary}.

\begin{remarks}\label{rem:7}
By Theorem~\ref{Thm:EJS-LB} and Proposition~\ref{Prop:BSC}, 
\begin{IEEEeqnarray}{rCl}
\mathbb{E}[\taute]\le  \frac{\log M + \log\log\frac{M}{\epsilon} }{C} +\frac{\log\frac{1}{\epsilon}+1}{C_1} + \frac{6(4C_2)^2}{C C_1},\nonumber\\
\end{IEEEeqnarray}
and thus the encoding rule described above  together with the decoding and stopping rules described in \eqref{ML_decoder} and \eqref{eq:deftau} 
achieves Burnashev's optimal asymptotic performance in~\eqref{FactUB}, see Corollary~\ref{cor1}.
\end{remarks}

In the following we present two algorithms that at each time $t=0,1,\ldots, \taute-1$ and for  given posterior vector $\boldsymbol{\rho}(t)$ implement encoding functions $\gamma$ satisfying~\eqref{eq:detbinarycond}.


\setlength{\algomargin}{2em}
\begin{algorithm}
 \SetAlgoLined
\SetEndCharOfAlgoLine{.}
$\delta=1$\;
\For{$n=1,\ldots,2^{M}$}{$v = \mathrm{dec2bin}(n,M)$ \quad \% binary 
representation of $n$ with $M$ digits\;
$z=(2 v - 1) \times [\rho_1(t),\rho_2(t),\ldots,\rho_M(t)]^\intercal$\; 
    \If{$z>0 \ {\&\&} \ z<\delta$}{
    		$\delta=z$\;
    		$\hat{v}=v$\;
   		} }
		\For{$i=1,\ldots, M$}{$\gamma(i)=\hat{v}_i$ \quad \% $\hat{v}_i$ denotes $i$-th bit of $\hat{v}$\;

}
 \caption{ }
 \label{AlgHM1}
\end{algorithm}

\setlength{\algomargin}{2em}
\begin{algorithm}
 \SetAlgoLined
\SetEndCharOfAlgoLine{.}
 $S_0=\{1,2,...,M\}$ and $S_1=\emptyset$\;
 $r_0=1, r_1=0, \rho_\tmin=0$, and $\delta=1$\;
 \While{$\rho_\tmin < \delta$}{
 		$k=\argmin_{i\in S_0} \rho_i(t)$\;
    $S_0=S_0 - \{k\}$ and $S_1=S_1 \cup \{k\}$\;
 	  $r_0=r_0-\rho_k(t)$ and $r_1=r_1+\rho_k(t)$\;
    \If{$r_0 < r_1$}{
    		Swap $S_0$ and $S_1$\;
    		Swap $r_0$ and $r_1$\;
   		}
    $\delta=r_0-r_1$\;
    $\rho_\tmin=\min_{i\in S_0} \rho_i(t)$\;
 	}  
\For{$i=1,\ldots, M$}{$\gamma(i)= \begin{cases} 0 \textnormal{ if } i\in S_0\\
1 \textnormal{ if } i\in S_1
\end{cases}$\;
}
 \caption{ }
 \label{AlgHM2}
\end{algorithm}

\begin{proposition} \label{alg} Both Algorithms~\ref{AlgHM1} and \ref{AlgHM2} satisfy condition~\eqref{eq:detbinarycond}. 
Algorithm~\ref{AlgHM1} has computational complexity of order~$O(2^M)$ for each {encoding} step while Algorithm~\ref{AlgHM2} has complexity of order~$O(M^2)$.\footnote{The computational complexity of Algorithm~\ref{AlgHM1} is of the same order as that of MaxEJS which in each step requires to find an encoding function (among $2^M$ choices) that maximizes the EJS divergence between the conditional output distributions.
However, implementation of Algorithm~\ref{AlgHM1} is simpler since it only requires linear operations instead of computing the EJS divergence (which can be computationally intensive, especially for channels with large output alphabet set).
We should point out that both Algorithms~\ref{AlgHM1} and~\ref{AlgHM2} have high computational complexity and are not suitable for practical implementation.} 
\end{proposition}
The proof is given in Appendix~\ref{app:new}.

\begin{remarks}
In contrast to the previous one-phase schemes in \cite{Horstein63, Burnashev74, Shayevitz11}, the encoding processes described by Algorithms~\ref{AlgHM1} and \ref{AlgHM2} here are completely deterministic. 
By insisting on a deterministic encoding, we can match our scheme's inputs only \emph{approximately} 
to the capacity-achieving input distribution of $(1/2, 1/2)$. 
On the other hand, the proposed deterministic schemes 
are such that once a particular message's posterior passes a certain threshold, the transmitter assigns this message exclusively to one of the two inputs. 
This is critical to achieve the optimal error exponent $E_\tmin=C_1$.
\end{remarks}

\begin{remarks}
As it is shown in Appendix~\ref{app:proofbinary} and \ref{app:new} (see also \cite{ITW2012}), proofs of Propositions~\ref{Prop:BSC} and~\ref{alg} continue to hold for those binary-input channels 
with uniform capacity-achieving input distribution $\pi_0^\star=\pi_1^\star=1/2$ where for ease of notation we
assume that $C_1=D(P_0\|P_1)$. 
This class of channels  includes the class of channels for which (\ref{eq:channelcond}) holds, 
for example the binary symmetric channel~(BSC) with cross-over 
probability $p\in(0,1/2)$, as well as the non-symmetric channel in Fig.~\ref{fig:2in3out} for $\eta \in(0,1/2)$.
\begin{figure}[htp!]
\centering
\includegraphics[width=0.4\textwidth]{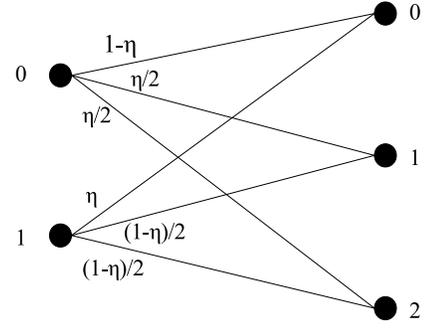}
\caption{Example of a non-symmetric (binary-input ternary-output) channel with capacity-achieving input distribution 
$\pi_0^\star=\pi_1^\star=1/2$.}
\label{fig:2in3out}
\end{figure}
\end{remarks}

\begin{remarks} 
The results in Proposition~\ref{Prop:BSC} and Remark~\ref{rem:7} above can also be extended to the case of $K$-ary symmetric channel with alphabet sets $\mathcal{X}=\mathcal{Y}=\{0,1,\ldots,K-1\}$
and transition probabilities of the form
\begin{align*} 
P(Y=y|X=x) = \begin{cases}  1-p &\mbox{if }  x=y\\
\frac{p}{K-1} &\mbox{if }  x\neq y
 \end{cases}
\end{align*}
where $p \in (0,\frac{K-1}{K})$.
Consider a coding scheme that at each time $t$ prior to the stopping time 
chooses the encoding function $\gamma$ in a way that if for any $x,x'\in\mathcal{X}$,
$$\sum_{j\in\Omega \colon \gamma(j)=x} \rho_j(t) \ge \max\bigg\{\frac{1}{K}, \sum_{j\in\Omega \colon \gamma(j)=x'} \rho_j(t)\bigg\},$$
then for all $i\in\{j\in\Omega \colon \gamma(j)=x\}$,
$$\sum_{j\in\Omega \colon \gamma(j)=x} \rho_j(t) - \sum_{j\in\Omega \colon \gamma(j)=x'} \rho_j(t) \leq \rho_i(t).$$
This coding scheme together with the decoding and stopping rules 
described in \eqref{ML_decoder} and \eqref{eq:deftau}
achieves Burnashev's optimal asymptotic performance in~\eqref{FactUB} for the $K$-ary symmetric channel.
\end{remarks}


\section{Reliability Function} \label{sec:rate-reliability}

Let a variable-length coding scheme $\fc$ be given that for each positive integer $\ell$ can transmit one out of $M_{\fc_\ell}$ equiprobable messages at a probability $\Pe_{\fc_\ell}$ and with an expected stopping time $\mathbb{E}_{\fc_\ell}[\tau]$.
If for any small numbers $\delta>0$, $0\leq \epsilon <1$ and all sufficiently large $\ell$  the following three conditions 
\begin{subequations}\label{eq:Ercond}
\begin{IEEEeqnarray}{rCl}
\Pe_{\fc_\ell}  & \leq & \epsilon \label{eq:errorell} \\
M_{\fc_\ell} & \geq  & 2^{\ell (R-\delta)}\label{eq:rateell}\\
\mathbb{E}_{\fc_\ell}[\tau] &  \leq & \ell,\label{eq:stopell}
\end{IEEEeqnarray}
\end{subequations}
hold for some positive real number $R$, then we say that the scheme $\fc$ achieves (information) rate $R$.\footnote{It would be more precise to talk about \emph{sequence of schemes} $\{\fc_\ell\}_{\ell\in\mathbb{Z}^+}$, where each $\fc_\ell$ is the general scheme $\fc$ specialized to the message size $M_{\fc_\ell}$. However, this would make the notation overcomplicated.}

If $\fc$ satisfies~\eqref{eq:rateell} and \eqref{eq:stopell} but instead of~\eqref{eq:errorell} it satisfies a stronger condition on 
exponential decay
\begin{equation}
\Pe_{\fc_\ell} \leq 2^{-\ell (E-\delta)}\label{eq:errorell-exp}
\end{equation}
for some positive real number $E$, then we say that the scheme $\fc$ achieves error exponent $E$ at rate~$R$.

 The capacity of a DMC is defined as the largest rate $R$ that is achievable over this channel; it is equal to the Shannon capacity $C$ as defined in~\eqref{eq:C} \cite[p.~184]{CoverBook}. For a given rate $R$ below capacity, the reliability function $E(R)$ is defined as the maximum achievable error exponent at rate~$R$.
By Burnashev's lower bound in~\eqref{FactLB}, we have the following lemma:

\begin{lemma}
\label{converse}
No coding scheme can achieve diminishing error probability at rates higher than $C$.
Furthermore, 
\begin{align}
E(R) \le C_1 \bigg(1-\frac{R}{C}\bigg),  \ R\in (0, C).
\end{align}
\end{lemma}

\begin{IEEEproof}[Proof of Lemma~\ref{converse}]
Let $\fc$ be a coding scheme that for each $\ell\in\mathbb{Z}^+$ and for a message size $M_{\fc_\ell}$ satisfies~\eqref{eq:Ercond} for a rate $R>0$. 

By~\eqref{FactLB} and \eqref{eq:Ercond},
for each sufficiently large integer $\ell$:
\begin{align}
\ell \ge \mathbb{E}_{\fc_\ell} [\tau] &\ge 
\bigg(\frac{\log M_{\fc_\ell}}{C} + \frac{\log(1/\Pe_{\fc_\ell}) }{C_1}\bigg)(1-o(1)) \nonumber \\ 
& \geq \bigg( \frac{ R\ell }{C} + \frac{\log(1/\Pe_{\fc_\ell}) }{C_1} \bigg)(1-o(1)). \label{RlessC}
\end{align}
In other words, 
\begin{align}
\label{RlessCv2}
\nonumber
C & \geq \bigg( R + \frac{C}{C_1}\cdot\frac{\log(1/\Pe_{\fc_\ell}) }{\ell} \bigg)(1-o(1))\\
& \geq R (1-o(1))
\end{align}
where the last inequality holds because $\log\frac{1}{\Pe_{\fc_\ell}}\geq 0$.
Since $o(1)\to 0$ as $\Pe_{\fc_\ell}\to 0$, we obtain from \eqref{RlessCv2} that $R \le C$. 
This implies that no coding scheme can achieve 
diminishing error probability at rates higher than $C$.

Next we characterize an upper bound on the optimal reliability function $E(R)$.
Let $\fc$ be a coding scheme that for each $\ell\in\mathbb{Z}^+$ and for a message size $M_{\fc_\ell}$ satisfies~\eqref{eq:rateell}, \eqref{eq:stopell}, and~\eqref{eq:errorell-exp} for $E, R>0$.
By~\eqref{FactLB},  \eqref{eq:rateell}, and \eqref{eq:errorell-exp}, for each sufficiently large integer $\ell$:
\begin{align}
\ell \ge \mathbb{E}_{\fc_\ell} [\tau] &\ge 
\bigg(\frac{\log M_{\fc_\ell}}{C} + \frac{\log(1/\Pe_{\fc_\ell})}{C_1}\bigg)(1-o(1)) \nonumber \\ 
& \geq \bigg(\frac{ R\ell }{C} + \frac{E \ell }{C_1}\bigg) (1-o(1)).  \label{achieve}
\end{align}
In other words, 
\begin{align}
1 \ge \bigg(\frac{R}{C} + \frac{E}{C_1}\bigg) (1-o(1)). 
\end{align}
Since $o(1)\to 0$ as $\ell \to \infty$, 
we obtain that $\frac{R}{C} + \frac{E}{C_1} \le 1$.
The desired inequality follows:
\begin{align}\label{eq:in1}
E \leq C_1 \left(1- \frac{R}{C}\right).
\end{align}
\end{IEEEproof}

On the other hand, we have the following achievable bound on the reliability function:

\begin{lemma}
\label{achievability}
Suppose that we have a coding scheme $\fc$ that for each message size $M>0$ and each positive $\epsilon>0$,  satisfies $\Pe_\fc\le \epsilon$ with expected stopping time 
\begin{equation}\label{eq:labelEbb}
\mathbb{E}_\fc[\tau]\le \left(\frac{\log M}{R_\tmin} +\frac{\log\frac{1}{\epsilon}}{E_\tmin}\right)(1+o(1))
\end{equation}
for some positive integers $E_\tmin$ and $R_\tmin$. 
Then, the scheme $\fc$ can achieve any rate $R\in[0,R_\tmin]$ with error exponent $E$, if
\begin{align}
\label{E2reliab}
E \le E_\tmin \left(1-\frac{R}{R_\tmin}\right).
\end{align}
\end{lemma}
Thus, if a scheme $\fc$ satisfies~\eqref{eq:labelEbb} for $R_\tmin=C$ and $E_\tmin=C_1$, then this scheme achieves Burnashev's reliability function. 
\begin{IEEEproof}[Proof of Lemma~\ref{achievability}]
Fix a small $\delta>0$, a positive rate $R<R_{\tmin}$ and a positive error exponent $E$ satisfying~\eqref{E2reliab}. Define for each  $\ell\in\mathbb{Z}^+$, the small number $\epsilon_\ell \triangleq 2^{-\ell(E-\delta)}$ and the message size $M_\ell \triangleq 2^{\ell (R-\delta)}$. By assumption, for each $\ell\in\mathbb{Z}^+$, our coding scheme $\fc$ attains a probability of error $\Pe_{\fc_\ell} \leq \epsilon_\ell$ at an expected stopping time $\mathbb{E}_{\fc_\ell}[\tau_{\epsilon_\ell}]$ that is upper bounded as:
\begin{IEEEeqnarray}{rCl}
\mathbb{E}_{\fc_\ell}[\tau_{\epsilon_\ell}] & \leq & \ell \left(\frac{ R-\delta}{R_\tmin} +\frac{ E-\delta}{E_\tmin}\right)(1+o(1)) \nonumber \\
& \leq  & \ell\left(1- \frac{\delta}{R_\tmin}-\frac{\delta}{E_\tmin} \right)(1+o(1)).
\end{IEEEeqnarray}
Since $\delta>0$ and since $o(1)\to 0$ as $\ell \to \infty$, we obtain that for sufficiently large $\ell$, 
\begin{IEEEeqnarray}{rCl}
\mathbb{E}_{\fc_\ell}[\tau_{\epsilon_\ell}]\leq \ell.
\end{IEEEeqnarray}
Combined with our assumptions that $\Pe_{\fc_\ell}\le 2^{-\ell(E-\delta)}$ and  $M_\ell \triangleq 2^{\ell (R-\delta)}$, this concludes the proof.
\end{IEEEproof}

Corollary~\ref{cor1} combined with Lemma~\ref{achievability} provides the following:

\begin{corollary}[Rate--Reliability Test]
\label{Cor:Test}
Consider a DMC with $C>0$ and $C_1 < \infty$ and a variable-length coding scheme~$\fc$.
If---irrespective of the size of the message set $M$---for any time~$t$ prior to the stopping time and for any posterior vector $\boldsymbol{\rho}(t)$ over the messages, the scheme selects (a possibly random) encoding function $\Gamma^\fc$ such that
\begin{subequations}
\begin{align}
\label{CapCond}
EJS(\boldsymbol{\rho}(t),\Gamma^\fc) \geq C, 
\end{align}
then it achieves the capacity $C$ of the channel. 
Furthermore, if also,
\begin{align}
\label{TestCond}
EJS(\boldsymbol{\rho}(t),\Gamma^\fc) \geq \tilde{\rho} C_1 \qquad \text{if} \ \max_{i\in\Omega} \rho_i(t) \geq \tilde{\rho}, 
\end{align}
then the scheme also achieves the optimal error exponent $E(R)$ of the channel. 
\end{subequations}
\end{corollary}

The above corollary implies that all coding schemes described in Sections~\ref{Sec:Main} and~\ref{Sec:Sym} achieve 
the capacity $C$ of the corresponding channels. 
Furthermore, the MaxEJS coding scheme and the simple coding scheme for the symmetric binary-input channel discussed in Section~\ref{sec:BinaryOpt} achieve Burnashev's reliability function $E(R)$.


\section*{Acknowledgements}
We like to acknowledge Hessam Mahdavifar for suggesting Algorithm 2 and Bar{\i}\c{s} Nakibo\u{g}lu for noting the validity of Proposition~\ref{Prop:BSC} for the class of $K$-ary symmetric channels given in Remark~10. 
We would also like to sincerely thank Igal Sason for his careful review of our material on arXiv as well his numerous suggestions how to improve the readability of our paper.
Additionally, we would like to thank Todd Coleman, Young-Han Kim, Yury Polyanskiy, Maxim Raginsky, Sergio Verd\'u, and Yihong Wu for valuable discussions and suggestions. 
Last but not least, we are grateful to the Associate Editor and the reviewers for their constructive comments.

\appendices

\section{Proof of Lemma~\ref{lem:propEJS}}\label{app:EJSproperties}
Property~\ref{item:EJS-JS} is proved as follows:
\begin{align*}
\lefteqn{JS(\boldsymbol{\rho};P_1,\ldots,P_M)}\\
&= \sum_{i=1}^M \rho_i D\bigg(P_i\|\sum_{j=1}^M \rho_j P_j\bigg)\\
&= \sum_{i=1}^M \rho_i D\bigg(P_i\| \rho_i P_i + (1-\rho_i)\sum_{j\neq i} \frac{\rho_j}{1-\rho_i} P_j\bigg)\\
&\stackrel{(a)}{\le} \sum_{i=1}^M \bigg [ \rho_i^2 D(P_i\| P_i) + \rho_i(1-\rho_i)D\bigg(P_i\|\sum_{j\neq i} \frac{\rho_j}{1-\rho_i} P_j\bigg) \bigg ]\\
&= EJS(\boldsymbol{\rho};P_1,\ldots,P_M) - \sum_{i=1}^M \rho_i^2 D\bigg(P_i\|\sum_{j\neq i} \frac{\rho_j}{1-\rho_i} P_j\bigg)\\
&\stackrel{(b)}{\le} EJS(\boldsymbol{\rho};P_1,\ldots,P_M)
\end{align*}
where $(a)$ and $(b)$ follow respectively because
 KL divergence is convex (in both arguments) and non-negative.

The proof of property~\ref{item:EJS-U} is provided next. 
\begin{align*}
\lefteqn{EJS(\boldsymbol{\rho};P_1,\ldots,P_M)}\\
&= \sum_{i=1}^M \rho_i D\bigg(P_i\|\sum_{j\neq i} \frac{\rho_j}{1-\rho_i} P_j\bigg)\\
&= \sum_{i=1}^M \rho_i \sum_{y\in\mathcal{Y}} P_i(y) \log\frac{P_i(y)}{\sum_{j\neq i} \frac{\rho_j}{1-\rho_i} P_j(y)}\\
&= \sum_{i=1}^M \rho_i \log\frac{1-\rho_i}{\rho_i} 
+ \sum_{i=1}^M \sum_{y\in\mathcal{Y}} \rho_i P_i(y) \log\frac{\rho_i P_i(y)}{\sum_{j\neq i} \rho_j P_j(y)}\\
&= U(\boldsymbol{\rho}) +  
\sum_{y\in\mathcal{Y}} P_{\boldsymbol{\rho}}(y) \sum_{i=1}^M \frac{\rho_i P_i(y)}{P_{\boldsymbol{\rho}}(y)} \log\frac{\frac{\rho_i P_i(y)}{P_{\boldsymbol{\rho}}(y)}}{1-\frac{\rho_i P_i(y)}{P_{\boldsymbol{\rho}}(y)}}\\
&= U(\boldsymbol{\rho}) - \sum_{y\in\mathcal{Y}} P_{\boldsymbol{\rho}}(y) U\Big(\Big[\frac{\rho_1 P_1(y)}{P_{\boldsymbol{\rho}}(y)},\ldots,\frac{\rho_M P_M(y)}{P_{\boldsymbol{\rho}}(y)}\Big]\Big).
\end{align*}

Property~\ref{item:EJS-cnvx} is proved as follows. 

Let $P_1,P_2,\ldots,P_M$ and $Q_1,Q_2,\ldots,Q_M$ be two sets of distributions.
For any $\lambda\in[0,1]$ and $\bar{\lambda}=1-\lambda$,
\begin{align*}
\lefteqn{EJS(\boldsymbol{\rho};\lambda P_1 + \bar{\lambda} Q_1,\ldots,\lambda P_M + \bar{\lambda} Q_M)}\\
&= \sum_{i=1}^M \rho_i D\Big(\lambda P_i + \bar{\lambda} Q_i\|\sum_{j\neq i} \frac{\rho_j}{1-\rho_i} \lambda P_j + \sum_{j\neq i} \frac{\rho_j}{1-\rho_i} \bar{\lambda} Q_j \Big )\\
&\stackrel{(a)}{\le} \sum_{i=1}^M \rho_i \Big [\lambda D\Big(P_i\|\sum_{j\neq i} \frac{\rho_j}{1-\rho_i} P_j\Big)\\
&\hspace{1.75in} + \bar{\lambda} D\Big(Q_i\|\sum_{j\neq i} \frac{\rho_j}{1-\rho_i} Q_j\Big) \Big ]\\
&= \lambda EJS(\boldsymbol{\rho};P_1,\ldots,P_M) + \bar{\lambda} EJS(\boldsymbol{\rho};Q_1,\ldots,Q_M)
\end{align*}
where $(a)$ follows because KL divergence is convex in both arguments.


\section{Proof of Theorem~\ref{Thm:EJS-LB}}
\label{App:Thm}

Let $\mathcal{F}(t)$ denote the history of the receiver's knowledge up to time $t$, i.e., 
$\mathcal{F}(t)=\sigma \{ Y^{t-1} \}$. 
Moreover, for each time $t=0,1,\ldots,\tau$, define
$$\tilde{U}(t):=\sum_{i=1}^M \rho_i(t) \log \frac{\rho_i(t)}{1-\rho_i(t)} - \log\frac{\tilde{\rho}}{1-\tilde{\rho}}$$
where recall that we defined $\tilde{\rho}=1-\frac{1}{1+\max\{\log M, \log\frac{1}{\epsilon}\}}$.
(For $M\ge2$ and $\epsilon\le 1$ which is the region of interest for these parameters,
 $\tilde{\rho}\ge\frac{1}{2}$.)
%

Notice that for all $i \in \Omega$ and given the observation $Y^{t-1}=y^{t-1}$, upon observing the new sample $y_t$, the belief state evolves as 
\begin{align}
\label{EvolveState}
\nonumber
\lefteqn{\rho_i(t+1)}\\ \nonumber
&= \Pr(\Theta=i|Y^t)\\ \nonumber
&= \frac{\Pr(\Theta=i,Y^t)}{\Pr(Y^t)}\\ \nonumber
&= \frac{\Pr(\Theta=i|Y^{t-1}=y^{t-1}) \Pr(Y_t=y_t|\Theta=i,Y^{t-1}=y^{t-1})}{\sum_{j=1}^M \Pr(\Theta=j,Y^t)}\\ \nonumber
&= \frac{\rho_i(t) \Pr(Y_t=y_t|\Theta=i,Y^{t-1}=y^{t-1})}{\sum \limits_{j=1}^M \rho_j(t) \Pr(Y_t=y_t|\Theta=j,Y^{t-1}=y^{t-1})}\\ \nonumber
&= \frac{\rho_i(t) \Pr(Y_t=y_t|X_t=\gamma_{y^{t-1}}(i))}{\sum \limits_{j=1}^M \rho_j(t) \Pr(Y_t=y_t|X_t=\gamma_{y^{t-1}}(j))}\\
&= \frac{\rho_i(t) P_{\gamma_{y^{t-1}}(i)}(y_t)}{\sum \limits_{j=1}^M \rho_j(t) P_{\gamma_{y^{t-1}}(j)}(y_t)}.
\end{align}

Furthermore,
\begin{align}
\label{ProbYt}
\nonumber
\lefteqn{\Pr(Y_t=y|Y^{t-1}=y^{t-1})}\\ \nonumber
&= \sum_{j=1}^M \Pr(Y_t=y, \Theta=j|Y^{t-1}=y^{t-1})\\ \nonumber
&= \sum_{j=1}^M \Pr(\Theta=j|Y^{t-1}=y^{t-1})\times \\ \nonumber 
&\qquad \Pr(Y_t=y|\Theta=j, Y^{t-1}=y^{t-1})\\ \nonumber
&= \sum_{j=1}^M \rho_j(t) \Pr(Y_t=y|X_t=\gamma_{y^{t-1}}(j))\\
&= \sum_{j=1}^M \rho_j(t) P_{\gamma_{y^{t-1}}(j)}(y).
\end{align}

%
From \eqref{EvolveState} and \eqref{ProbYt}, under a (possibly randomized) coding scheme $\fc$,
%
\begin{align*}
\lefteqn{\mathbb{E}_\fc \left [\sum_{i=1}^M \rho_i(t+1) \log\frac{\rho_i(t+1)}{1-\rho_i(t+1)} \bigg| \mathcal{F}(t) \right ]} \\
&= \sum_{\gamma\in\mathcal{E}} \Pr(\Gamma^\fc=\gamma|Y^{t-1})
\sum_{y \in \mathcal{Y}} \left( \sum_{j=1}^M \rho_j(t) P_{\gamma(j)}(y) \right) \times \\ 
&\qquad \Bigg[ \sum_{i=1}^M \frac{\rho_i(t) P_{\gamma(i)}(y)}{\sum \limits_{j=1}^M \rho_j(t) P_{\gamma(j)}(y)}
\log\frac{\rho_i(t)P_{\gamma(i)}(y)}{\sum \limits_{j \neq i} \rho_j(t)P_{\gamma(j)}(y)}\Bigg] \\
&= \sum_{\gamma\in\mathcal{E}} \Pr(\Gamma^\fc=\gamma|Y^{t-1}) \times \\
&\qquad \sum_{y \in \mathcal{Y}} \sum_{i=1}^M \rho_i(t) P_{\gamma(i)}(y) 
\log\frac{\rho_i(t)P_{\gamma(i)}(y)}{\sum \limits_{j \neq i} \rho_j(t)P_{\gamma(j)}(y)}\\
&= \sum_{i=1}^M \rho_i(t)\log\frac{\rho_i(t)}{1-\rho_i(t)} + \sum_{\gamma\in\mathcal{E}} \Pr(\Gamma^\fc=\gamma|Y^{t-1})\times \\  
&\qquad \sum_{i=1}^M \rho_i(t) D\bigg(P_{\gamma(i)} \Big\| \sum_{j \neq i} \frac{\rho_j(t)}{1-\rho_i(t)} P_{\gamma(j)} \bigg)\\
&= \sum_{i=1}^M \rho_i(t)\log\frac{\rho_i(t)}{1-\rho_i(t)} + EJS(\boldsymbol{\rho}(t),\Gamma^\fc)
\end{align*}
%
%
which implies that
%
\begin{align}
\label{jumps}
\mathbb{E}_\fc \left [\tilde{U}(t+1) | \mathcal{F}(t) \right ] = \tilde{U}(t) + EJS(\boldsymbol{\rho}(t),\Gamma^\fc).
\end{align}

Note that if $\rho_i(t) < \tilde{\rho}$, $\forall i\in\Omega$, then $\tilde{U}(t)<0$.
Therefore, $\tilde{U}(t)\ge 0$ implies that $\exists i\in\Omega$ such that $\rho_i(t)\ge \tilde{\rho}$.
From~\eqref{jumps} and 
condition~\eqref{CondTestThm} of Theorem~\ref{Thm:EJS-LB}, the sequence $\{\tilde{U}(t)\}_{t=0}^{\tau}$ satisfies
\begin{align} 
\label{lowerbound-overall}
\mathbb{E}_\fc \left [\tilde{U}(t+1)  | \mathcal{F}(t) \right ] \geq \begin{cases}  \tilde{U}(t) + R_\tmin &\mbox{if }  \tilde{U}(t) < 0\\
 \tilde{U}(t) + \tilde{\rho} E_\tmin &\mbox{if } \tilde{U}(t) \ge 0
 \end{cases} . 
\end{align}
The sequence $\{\tilde{U}(t)\}_{t=0}^{\tau}$ forms a submartingale with respect to the filtration $\{\mathcal{F}(t)\}$.
Furthermore, from Lemma~\ref{drift:U} below,
\begin{align} 
\label{boundedjumps}
|\tilde{U}(t+1)-\tilde{U}(t)| \leq 4 C_2 \hspace{0.075 in} {\mbox{if}} \hspace{0.075 in} \max\{\tilde{U}(t),\tilde{U}(t+1)\} \ge 0.
\end{align}

Note that if $\rho_i(t)< 1-\epsilon$ for all $i\in\Omega$, then
$$\tilde{U}(t)< \sum_{i=1}^M \rho_i(t) \log \frac{1-\epsilon}{\epsilon} - \log\frac{\tilde{\rho}}{1-\tilde{\rho}} \le \log\frac{1-\epsilon}{\epsilon}.$$
In other words, if $\tilde{U}(t)\ge \log\frac{1}{\epsilon}$, then there is an $i\in\Omega$ for which
$\rho_i(t)\ge 1-\epsilon$.
Let $\upsilon:=\min\{t: \tilde{U}(t)\ge \log\frac{1}{\epsilon}\}$.
Note that by construction, $\taute \leq \upsilon$.
Appealing to Lemma~\ref{Lem:Martingale} at the end of this section, we obtain
\begin{IEEEeqnarray}{rCl}
\nonumber
\lefteqn{\mathbb{E}_\fc [\taute] \le \mathbb{E}_\fc [\upsilon]} \\ \nonumber
&\le& \frac{\log\frac{1}{\epsilon}-\tilde{U}(0)}{\tilde{\rho} E_\tmin}
+ \tilde{U}(0) {\bf{1}}_{\{\tilde{U}(0)< 0\}} 
\bigg(\frac{1}{\tilde{\rho} E_\tmin} - \frac{1}{R_\tmin} \bigg)\\ \nonumber
& & + \frac{3(4C_2)^2}{\tilde{\rho}R_\tmin E_\tmin}\\ \nonumber 
&\le& \frac{\log\frac{1}{\epsilon}}{\tilde{\rho} E_\tmin}
+ \frac{-\tilde{U}(0)}{R_\tmin} {\bf{1}}_{\{\tilde{U}(0)< 0\}} + \frac{6(4C_2)^2}{R_\tmin E_\tmin}\\
\nonumber 
&= &\frac{\log\frac{1}{\epsilon}}{\tilde{\rho} E_\tmin}
+ \frac{\sum \limits_{i=1}^M \rho_i(0)\log\frac{1-\rho_i(0)}{\rho_i(0)}+\log\frac{\tilde{\rho}}{1-\tilde{\rho}}}{R_\tmin} {\bf{1}}_{\{\tilde{U}(0)< 0\}}\\ \nonumber
& & + \frac{6(4C_2)^2}{R_\tmin E_\tmin}\\ \nonumber
&\le& \frac{\log\frac{1}{\epsilon}}{\tilde{\rho} E_\tmin}
+ \frac{H(\boldsymbol{\rho}(0))+\log\frac{\tilde{\rho}}{1-\tilde{\rho}}}{R_\tmin} + \frac{6(4C_2)^2}{R_\tmin E_\tmin}\\
\nonumber
&\le& \frac{\log\frac{1}{\epsilon}}{E_\tmin}\Big(1+\frac{1}{\max\{\log M,\log\frac{1}{\epsilon}\}}\Big)
+ \frac{H(\boldsymbol{\rho}(0))+\log\log\frac{M}{\epsilon}}{R_\tmin}\\ \nonumber
&&  + \frac{6(4C_2)^2}{R_\tmin E_\tmin}\\ 
&\le&\frac{H(\boldsymbol{\rho}(0))+ \log\log\frac{M}{\epsilon}}{R_\tmin} + \frac{\log\frac{1}{\epsilon}+1}{E_\tmin}
+  \frac{6(4C_2)^2}{R_\tmin E_\tmin}\label{UB01}.
\end{IEEEeqnarray}

%

\begin{lemma}
\label{drift:U}
If $\max \{\tilde{U}(t),\tilde{U}(t+1)\} \ge 0$, then 
\begin{align*}
\left|\tilde{U}(t+1)-\tilde{U}(t)\right| \leq 4 C_2.
\end{align*}
\end{lemma}

\begin{IEEEproof}
We first consider the case $\tilde{U}(t)\ge 0$. 
Note that if $\rho_i(t) < \tilde{\rho}$, $\forall i\in\Omega$, then $\tilde{U}(t)<0$.
Therefore, $\tilde{U}(t)\ge 0$ implies that $\exists i\in\Omega$ such that $\rho_i(t)\ge \tilde{\rho}$.
Without loss of generality assume $\rho_1(t)\ge \tilde{\rho}$. We obtain,
\begin{align*} 
\lefteqn{\left|\tilde{U}(t+1)-\tilde{U}(t)\right|}\\
&=\left|\sum_{i=1}^M \rho_i(t+1) \log\frac{\rho_i(t+1)}{1-\rho_i(t+1)} - \sum_{i=1}^M \rho_i(t) \log\frac{\rho_i(t)}{1-\rho_i(t)}\right|\\
&=\left|\sum_{i=1}^M \rho_i(t+1) \left(\log\frac{\rho_i(t+1)}{1-\rho_i(t+1)}-\log\frac{\rho_i(t)}{1-\rho_i(t)}\right) \right.\\
& \hspace*{1.09in} + \left. \sum_{i=1}^M \left(\rho_i(t+1)-\rho_i(t)\right) \log\frac{\rho_i(t)}{1-\rho_i(t)}\right|\\
&\le \max_{i \in \Omega} \left|\log\frac{\rho_i(t+1)}{1-\rho_i(t+1)}-\log\frac{\rho_i(t)}{1-\rho_i(t)}\right| \\
& \hspace*{1.09in} + \left|\sum_{i=1}^M \left(\rho_i(t+1)-\rho_i(t)\right) \log\frac{\rho_i(t)}{1-\rho_i(t)}\right|\\
&\stackrel{(a)}{\le} \log C_2 + \sum_{i=1}^M \left|\rho_i(t+1)-\rho_i(t)\right| \cdot \left|\log\frac{\rho_i(t)}{1-\rho_i(t)}\right|\\
&\stackrel{(b)}{\le} \log C_2 + C_2 \sum_{i=1}^M \rho_i(t)(1-\rho_i(t)) \left|\log\frac{\rho_i(t)}{1-\rho_i(t)}\right|\\
&\le \log C_2 + C_2 \rho_{1}(t)(1-\rho_{1}(t)) \log\frac{\rho_{1}(t)}{1-\rho_{1}(t)}\\
&\hspace*{.15in} + C_2 \sum_{i\neq {1}} \rho_i(t) \log\frac{1}{\rho_i(t)}\\
&\stackrel{(c)}{\le} \log C_2 + C_2 + C_2 \bigg(\sum_{i\neq {1}} \rho_i(t)\bigg) \log\frac{M-1}{\sum \limits_{i\neq {1}} \rho_i(t)}\\
&\le \log C_2 + C_2 + C_2 ( (1-\tilde{\rho}) \log (M-1) + 1)\\
&= \log C_2 + C_2 + C_2 \Big( \frac{\log (M-1)}{1+\max\{\log M,\log\frac{1}{\epsilon}\}}+1\Big)\\
&\le \log C_2 + 3 C_2\\
&\le 4 C_2
\end{align*}
where $(a)$ and $(b)$ follow respectively from Lemmas~\ref{drift:log} and \ref{drift:rho} below,
and $(c)$ follows from Jensen's inequality and the fact that
$$\Big |x(1-x) \log\frac{x}{1-x} \Big | \leq 1, \ \ \ x \in [0,1].$$ 

This completes the proof for the case $\tilde{U}(t)\ge 0$. 
The proof for the case $\tilde{U}(t+1)\ge 0$ is done by following the similar lines and interchanging time indices $(t)$ and $(t+1)$.
\end{IEEEproof}


\begin{lemma}
\label{drift:log}
For any $i \in \Omega$,
\begin{align*}
\left|\log\frac{\rho_i(t+1)}{1-\rho_i(t+1)}-\log\frac{\rho_i(t)}{1-\rho_i(t)}\right| \leq \log C_2.
\end{align*}
\end{lemma}

\begin{IEEEproof}
\begin{align*}
\lefteqn{\left|\log\frac{\rho_i(t+1)}{1-\rho_i(t+1)}-\log\frac{\rho_i(t)}{1-\rho_i(t)}\right|}\\
&=\left|\log\frac{P(Y=y_t|X=\gamma_{y^{t-1}}(i))}{\sum \limits_{j\neq i} \frac{\rho_j(t)}{1-\rho_i(t)} P(Y=y_t|X=\gamma_{y^{t-1}}(j))}\right|\\
&\le \max_{y \in \mathcal{Y}} \log\frac{\max_{x \in \mathcal{X}} P(Y=y|X=x)}{\min_{x \in \mathcal{X}} P(Y=y|X=x)} = \log C_2.
\end{align*}
\end{IEEEproof}

\begin{lemma}
\label{drift:rho}
For any $i \in \Omega$,
\begin{align*}
\lefteqn{\left|\rho_i(t+1)-\rho_i(t)\right|}\\ 
&\leq \min \left\{\rho_i(t)(1-\rho_i(t)), \rho_i(t+1)(1-\rho_i(t+1)) \right\}C_2.
\end{align*}
\end{lemma}

\begin{IEEEproof}
\begin{align}
\label{drift01}
\nonumber
\lefteqn{\left|\rho_i(t+1)-\rho_i(t)\right|}\\ \nonumber
&=\rho_i(t) \left|\frac{P(Y=y_t|X=\gamma_{y^{t-1}}(i))}{\sum \limits_{j=1}^M \rho_j(t) P(Y=y_t|X=\gamma_{y^{t-1}}(j))} - 1 \right|\\ \nonumber
&\le\rho_i(t) \left|\frac{(1-\rho_i(t)) \max_{x \in \mathcal{X}} P(Y=y_t|X=x)}{\sum \limits_{j=1}^M \rho_j(t) P(Y=y_t|X=\gamma_{y^{t-1}}(j))}\right|\\ \nonumber
&\le \rho_i(t)(1-\rho_i(t)) \max_{y \in \mathcal{Y}} \frac{\max_{x \in \mathcal{X}} P(Y=y|X=x)}{\min_{x \in \mathcal{X}} P(Y=y|X=x)}\\
&= \rho_i(t)(1-\rho_i(t)) C_2.
\end{align}

Similarly we can show that
\begin{align}
\label{drift02}
\nonumber
\lefteqn{\left|\rho_i(t+1)-\rho_i(t)\right|}\\ \nonumber
&=\rho_i(t+1) \left|1-\rho_i(t)-\frac{\sum \limits_{j\neq i} \rho_j(t) P(Y=y_t|X=\gamma_{y^{t-1}}(j))}{P(Y=y_t|X=\gamma_{y^{t-1}}(i))} \right|\\ \nonumber
&=\rho_i(t+1)(1-\rho_i(t+1)) \times \\ \nonumber 
& \qquad \left|\frac{1-\rho_i(t)}{1-\rho_i(t+1)} -\frac{\sum \limits_{j=1}^M \rho_j(t) P(Y=y_t|X=\gamma_{y^{t-1}}(j))}{P(Y=y_t|X=\gamma_{y^{t-1}}(i))} \right|\\ \nonumber
&\le \rho_i(t+1)(1-\rho_i(t+1)) \max_{y \in \mathcal{Y}} \frac{\max_{x \in \mathcal{X}} P(Y=y|X=x)}{\min_{x \in \mathcal{X}} P(Y=y|X=x)}\\
&= \rho_i(t+1)(1-\rho_i(t+1)) C_2.
\end{align}

Combining \eqref{drift01} and \eqref{drift02}, we have the assertion of the lemma.
\end{IEEEproof}

%
\begin{lemma}
\label{Lem:Martingale}
Assume that the sequence $\{\xi(t)\}$,  $t=0,1,2,\ldots$ forms a submartingale with respect to a filtration $\{\mathcal{F}(t)\}$.
Furthermore, assume there exist positive constants $K_1$, $K_2$, and $K_3$ such that
\begin{align*}
& \mathbb{E} [ \xi(t+1) | \mathcal{F}(t) ] \ge \xi(t) +K_1 \hspace{0.1 in} {\mbox{if}} \hspace{0.1 in} \xi(t) < 0,\\   
& \mathbb{E} [ \xi(t+1) | \mathcal{F}(t) ] \ge \xi(t) +K_2 \hspace{0.1 in} {\mbox{if}} \hspace{0.1 in} \xi(t) \ge 0,\\
& \left| \xi(t+1) - \xi(t) \right| \le K_3 \hspace{0.1 in} {\mbox{if}} \hspace{0.1 in} \max\{\xi(t+1),\xi(t)\} \ge 0.
\end{align*}
Consider the stopping time $\upsilon = \min \{ t: \xi(t) \ge B \}$, $B>0$.
Then we have the inequality
$$\mathbb{E} [\upsilon] \le  \frac{B - \xi(0)}{K_2} + \xi(0) {\bf{1}}_{\{\xi(0) <0\}} \left(\frac{1}{K_2} - \frac{1}{K_1} \right)+ 
\frac{3 K_3^2}{K_1 K_2}.$$ 
\end{lemma}
\begin{IEEEproof}
This lemma is a generalization of Lemma~1 in~\cite{Burnashev75}.
The proof is provided below.

Consider the sequence $\{\eta(t)\}$ defined as follows
\begin{align*} 
 \eta(t) = \begin{cases} -A + \frac{\xi(t)}{K_1} - t &\mbox{if }  \xi(t) < 0\\ 
 -A e^{-\alpha \xi(t)} + \frac{\xi(t)}{K_2} - t &\mbox{if } \xi(t) \ge 0
\end{cases}
\end{align*}
where $A=\left[\frac{3K_3^2}{K_2} \left(\frac{1}{K_1}-\frac{1}{K_2}\right)\right]^+$ and $\alpha=\frac{0.5 K_2}{K_3^2}$.

\begin{claim}
\label{eta_subm}
The sequence $\{\eta(t)\}$
forms a submartingale with respect to the filtration $\{\mathcal{F}(t)\}$.
\end{claim}

%
By Doob's Stopping Theorem,
\begin{align*}
\eta(0) &\le \mathbb{E}[\eta(\upsilon)]\\
&\le \mathbb{E}\left[\frac{\xi(\upsilon)}{K_2} - \upsilon \right]\\
 &= \frac{\mathbb{E}\left[\xi(\upsilon-1)\right]+\mathbb{E}\left[\xi(\upsilon)-\xi(\upsilon-1)\right]}{K_2} - \mathbb{E}[\upsilon]\\
 &\le \frac{B+K_3}{K_2} - \mathbb{E}[\upsilon].
\end{align*}

On the other hand, we have
\begin{align*}
\eta(0) &= \left(-A + \frac{\xi(0)}{K_1}\right) {\bf{1}}_{\{\xi(0) <0\}}\\
&\hspace*{.15in} + \left(-A e^{-\alpha \xi(0)} + \frac{\xi(0)}{K_2} \right) {\bf{1}}_{\{\xi(0) \ge 0\}}\\
&\ge -A + \frac{\xi(0)}{K_2} - \xi(0){\bf{1}}_{\{\xi(0) <0\}} \left(\frac{1}{K_2} - \frac{1}{K_1} \right).
\end{align*}

Combining the above inequalities, we obtain
\begin{align}
\nonumber
\lefteqn{\mathbb{E}[\upsilon]}\\ 
\nonumber
&\le \frac{B+K_3}{K_2} - \eta(0)\\
\nonumber
&\le\frac{B+K_3}{K_2}+A-\frac{\xi(0)}{K_2}+\xi(0){\bf{1}}_{\{\xi(0)<0\}}\left(\frac{1}{K_2}-\frac{1}{K_1} \right)\\
\nonumber
&= \frac{B-\xi(0)}{K_2}+\xi(0){\bf{1}}_{\{\xi(0)<0\}}\left(\frac{1}{K_2}-\frac{1}{K_1} \right)\\
\nonumber
&\hspace*{.15in} + \left[\frac{3K_3^2}{K_2} \left(\frac{1}{K_1}-\frac{1}{K_2}\right)\right]^+ + \frac{K_3}{K_2}\\
&\stackrel{(a)}{\le} \frac{B-\xi(0)}{K_2}+\xi(0){\bf{1}}_{\{\xi(0)<0\}}\left(\frac{1}{K_2}-\frac{1}{K_1} \right) + \frac{3K_3^2}{K_1 K_2}
\end{align}
where $(a)$ holds since by definition $K_1, K_2 \le K_3$ and hence, 
$\frac{K_3}{K_2}\le\min\left\{\frac{3K_3^2}{K_1 K_2},\frac{3K_3^2}{K_2^2}\right\}$.
\end{IEEEproof}

\vspace*{.1in}

\begin{IEEEproof}[Proof of Claim~\ref{eta_subm}]
We will show that $\mathbb{E}[\eta(t+1)|\mathcal{F}(t)]\ge\eta(t)$.
There are two cases:

{\bf{Case I. $\xi(t)<0$:}}

If $\xi(t+1)<0$, then
\begin{align}
\label{C1Neg}
\eta(t+1)=-A+\frac{\xi(t+1)}{K_1}-(t+1).
\end{align}
On the other hand, if $\xi(t+1)\ge 0$, then by the assumption of Lemma~\ref{Lem:Martingale},
$\xi(t+1)\le K_3$, and we have 
\begin{align}
\label{C1Pos}
\nonumber
\eta(t+1) &= -A e^{-\alpha \xi(t+1)}+\frac{\xi(t+1)}{K_2}-(t+1)\\
&\stackrel{(a)}{\ge} -A+\frac{\xi(t+1)}{K_1}-(t+1)
\end{align}
where $(a)$ follows from the fact that
1) if $K_1 \ge K_2$, then by definition $A=0$, and $\frac{x}{K_2} \ge \frac{x}{K_1}$ for $x\ge 0$; and
2) if $K_1 < K_2$, then 
$-A e^{-\alpha x}+\frac{x}{K_2}$ is concave in $x$,
$-A e^{-\alpha x}+\frac{x}{K_2}=-A+\frac{x}{K_1}$ for $x=0$, and for $x=K_3$
\begin{align}
\label{ConcavePointK3}
\nonumber
-A e^{-\alpha K_3}+\frac{K_3}{K_2} &\ge -A(1-\alpha K_3+\frac{1}{2}(\alpha K_3)^2)+\frac{K_3}{K_2}\\ \nonumber
&= -A + A \alpha K_3 (1-\frac{1}{4}\frac{K_2}{K_3})+\frac{K_3}{K_2}\\ \nonumber
&\ge -A + \frac{9}{8} K_3 \left(\frac{1}{K_1}-\frac{1}{K_2}\right) +\frac{K_3}{K_2}\\
&\ge -A+\frac{K_3}{K_1}.
\end{align}

Combining \eqref{C1Neg} and \eqref{C1Pos}, we obtain
\begin{align}
\label{C1exp}
\nonumber
\mathbb{E}[\eta(t+1)|\mathcal{F}(t)] &\ge \mathbb{E}[-A+\frac{\xi(t+1)}{K_1}-(t+1)|\mathcal{F}(t)]\\
\nonumber
&\ge -A+\frac{\xi(t)+K_1}{K_1}-(t+1)\\
&= -A+\frac{\xi(t)}{K_1}-t = \eta(t).
\end{align}

{\bf{Case II. $\xi(t)\ge 0$:}}

If $\xi(t+1)\ge 0$, then
\begin{align}
\label{C2Pos}
\eta(t+1) &= -A e^{-\alpha \xi(t+1)}+\frac{\xi(t+1)}{K_2}-(t+1).
\end{align}
On the other hand, if $\xi(t+1)< 0$, then we have
\begin{align}
\label{C2Neg}
\nonumber
\eta(t+1) &= -A+\frac{\xi(t+1)}{K_1}-(t+1)\\
&\stackrel{(a)}{\ge} -A e^{-\alpha \xi(t+1)}+\frac{\xi(t+1)}{K_2}-(t+1)
\end{align}
where $(a)$ follows from the fact that
1) if $K_1 \ge K_2$, then by definition $A=0$, and $\frac{x}{K_1} \ge \frac{x}{K_2}$ for $x< 0$; and
2) if $K_1 < K_2$, then 
$-A e^{-\alpha x}+\frac{x}{K_2}$ is concave in $x$,
$-A e^{-\alpha x}+\frac{x}{K_2}=-A+\frac{x}{K_1}$ for $x=0$, and for $x=K_3$ from \eqref{ConcavePointK3} we have
$-A e^{-\alpha K_3}+\frac{K_3}{K_2} \ge -A+\frac{K_3}{K_1}$. Note that if function $f$ is concave and $g$ is linear, $f(0)=g(0)$, and $f(b) \ge g(b)$ for some $b > 0$, then $f(x) \le g(x) $ for all $x\leq 0$.

Combining \eqref{C2Pos} and \eqref{C2Neg}, we obtain
\begin{align}
\label{C2exp}
\nonumber
\lefteqn{\mathbb{E}[\eta(t+1)|\mathcal{F}(t)]}\\
\nonumber
&\ge \mathbb{E}[-A e^{-\alpha \xi(t+1)}+\frac{\xi(t+1)}{K_2}-(t+1)|\mathcal{F}(t)]\\
\nonumber
&\ge \mathbb{E}[-A e^{-\alpha \xi(t+1)}|\mathcal{F}(t)] +\frac{\xi(t)+K_2}{K_2}-(t+1)\\
\nonumber
&= \mathbb{E}[-A e^{-\alpha \xi(t+1)}|\mathcal{F}(t)] +A e^{-\alpha \xi(t)}+\eta(t)\\
\nonumber
&= \eta(t)-A e^{-\alpha \xi(t)} \mathbb{E}[e^{-\alpha(\xi(t+1)-\xi(t))}-1|\mathcal{F}(t)]\\
\nonumber
&\stackrel{(a)}{\ge} \eta(t)-A e^{-\alpha \xi(t)} 
\mathbb{E}[-\alpha(\xi(t+1)-\xi(t))\\
\nonumber
&\hspace*{1.25in} + \frac{1}{2} \alpha^2(\xi(t+1)-\xi(t))^2 e^{\alpha K_3}|\mathcal{F}(t)]\\
\nonumber
\nonumber
&\ge \eta(t) + A \alpha e^{-\alpha \xi(t)} [K_2 - \frac{1}{2} \alpha K_3^2 e^{\alpha K_3}] \\
&\stackrel{(b)}{\ge} \eta(t)
\end{align}
where $(a)$ follows from the fact that for $|x|\le K$,
\begin{align*}
e^x &= 1 + \sum_{n=1}^\infty \frac{x^n}{n!}\\
&\le 1 + x + \frac{x^2}{2} \Big (1 + \frac{K}{3} + \frac{K^2}{12} + \ldots \Big )\\
&\le 1 + x + \frac{x^2}{2} e^{K};
\end{align*}
and $(b)$ holds since
\begin{align*} 
\frac{1}{2} \alpha K_3^2 e^{\alpha K_3} = \frac{1}{4} K_2 e^{\frac{0.5 K_2}{K_3}}\le \frac{e^{0.5}}{4} K_2 \le K_2.
\end{align*}
\end{IEEEproof}

\section{Proof of the Propositions}
\label{App:Props}

\subsection{Proof of Proposition~\ref{prop:PM}}\label{app:PM}
Fix a time instant $t$ and assume that $Y^{t-1}=y^{t-1}$. For ease of notation, in the following we drop the time index $t$ for $\rho_i(t)$ and simply write $\rho_i$.

Let 
\[
\lambda_{\gamma} := \Pr(\Gamma^{\textnormal{PM}}=\gamma|Y^{t-1}=y^{t-1}).
\]
Define for each $i\in\Omega$ and $x\in\mathcal{X}$:
\begin{IEEEeqnarray}{rCl}
\Lambda_{i,x} := \sum_{\gamma\colon\gamma(i)=x} \lambda_\gamma= \Pr(X=x| \Theta=i, Y^{t-1}=y^{t-1}) \IEEEeqnarraynumspace
\end{IEEEeqnarray}
and \begin{equation}
\hat \rho_{i,x} := \rho_i \Lambda_{i,x}= \Pr(X=x, \Theta=i| Y^{t-1}=y^{t-1}).
\end{equation}
Notice that for each $i,j\in\Omega$, $x,x'\in\mathcal{X}$, and for a fixed posterior distribution, 
the various messages are mapped into inputs of the channel independently of each other and hence,
\begin{IEEEeqnarray}{rCl}\label{eq:prod}
\sum_{ \gamma\colon \substack{\gamma(i)=x\\\gamma(j)=x'}} \lambda_\gamma = \Lambda_{i,x}\Lambda_{j,x'}.
\end{IEEEeqnarray}

Rearranging terms and using Jensen's inequality, we obtain
\begin{align}
\lefteqn{EJS(\boldsymbol{\rho}(t), \Gamma^\mathrm{PM})}  \nonumber \\
&= \sum_{\gamma \in \mathcal{E}} \lambda_\gamma \sum_{i=1}^M  \rho_i 
D\bigg(P_{\gamma(i)}\Big\|\sum_{j \neq i} \frac{\rho_j}{1-\rho_i} P_{\gamma(j)} \bigg) \nonumber \\
&=  \sum_{i=1}^{M} \rho_i \sum_{x\in \mathcal{X}}   \sum_{\gamma \colon \gamma(i)=x} \lambda_\gamma 
D\bigg(P_x\Big\| \sum_{j\neq i} \frac{\rho_j}{1-\rho_i}  P_{\gamma(j)} \bigg) \nonumber \\
& \geq \sum_{i=1}^M \sum_{x\in\mathcal{X}}  \rho_i  \Lambda_{i,x}   D\bigg({P}_x \Big\| \sum_{j\neq i} \frac{\rho_j}{1-\rho_i}\sum_{\gamma\colon \gamma(i)=x} \frac{\lambda_\gamma}{\Lambda_{i,x}} P_{\gamma(j)}\bigg) \nonumber \\
& = \sum_{i=1}^M \sum_{x\in\mathcal{X}}   \hat{\rho}_{i,x}   D\bigg({P}_x \Big\| \sum_{j\neq i} \frac{\rho_j}{1-\rho_i}\sum_{x'\in\mathcal{X}} \sum_{ \gamma\colon \substack{\gamma(i)=x\\\gamma(j)=x'}} 
 \frac{\lambda_\gamma}{\Lambda_{i,x}} P_{x'}\bigg) \nonumber \\
& \stackrel{(a)}{=}  \sum_{i=1}^M \sum_{x\in\mathcal{X}}  \hat{\rho}_{i,x}   D\bigg({P}_x \Big\|  \frac{\sum_{j\neq i} \sum_{x' \in \mathcal{X}} \rho_j \Lambda_{j,x'} P_{x'} }{1-\rho_i}\bigg)\nonumber \\
& =  \sum_{i=1}^M \sum_{x\in\mathcal{X}}  \hat{\rho}_{i,x}  D\bigg({P}_x \Big\| \frac{\sum_{x' \in \mathcal{X}} (\pi^\star_{x'} P_{x'} -  \hat{\rho}_{i,x'} {P}_{x'}) }{1-\rho_i}\bigg) \nonumber \\
& =   \sum_{i=1}^M \sum_{x\in\mathcal{X}}  \hat{\rho}_{i,x}  D\bigg({P}_x \Big\| \frac{\sum_{x' \in \mathcal{X}} (\pi^\star_{x'} P_{x'} -  \hat{\rho}_{i,x'} {P}_{x'}) }{1-\rho_i}\bigg)\nonumber \\
& \qquad + \sum_{i=1}^M \sum_{x\in\mathcal{X}} \hat{\rho}_{i,x}   \frac{\rho_i}{1-\rho_i} D\bigg({P}_x \Big\| \frac{\sum_{x'}\hat{\rho}_{i,x'} {P}_{x'} }{\rho_i}\bigg) \nonumber \\
& \qquad - \sum_{i=1}^M \sum_{x\in\mathcal{X}} \hat{\rho}_{i,x}   \frac{\rho_i}{1-\rho_i} D\bigg({P}_x \Big\| \frac{\sum_{x'}\hat{\rho}_{i,x'} {P}_{x'} }{\rho_i}\bigg) \nonumber \\
& \geq \sum_{i=1}^M \sum_{x\in\mathcal{X}} \frac{ \hat{\rho}_{i,x} }{1-\rho_i} D\bigg({P}_x \Big\| \sum_{x' \in \mathcal{X}} \pi^\star_{x'} P_{x'} \bigg)\nonumber\\
& \qquad - \sum_{i=1}^M \frac{\rho_i^2}{1-\rho_i}  \sum_{x\in\mathcal{X}}\Lambda_{i,x}   D\bigg({P}_x \Big\|  \sum_{x' \in \mathcal{X}} \Lambda_{i,x'} {P}_{x'} \bigg) \nonumber \\
& \stackrel{(b)}{\geq}  \sum_{i=1}^M \sum_{x\in\mathcal{X}} \frac{\hat{\rho}_{i,x}}{1-\rho_i}C - \sum_{i=1}^M    \frac{\rho_i^2 }{1-\rho_i} C \nonumber \\
& =   \sum_{i=1}^M \frac{\rho_i}{1-\rho_i} C -  \sum_{i=1}^M \frac{\rho_i^2}{1-\rho_i}  C \nonumber \\
& =C
\end{align} 
where $(a)$ follows from \eqref{eq:prod}; and inequality $(b)$ follows from Fact~\ref{Gallager} and that $$\sum_{x\in\mathcal{X}}\Lambda_{i,x} D\Big({P}_x \big\|  \sum_{x' \in \mathcal{X}} \Lambda_{i,x'} {P}_{x'} \Big)$$ is the mutual information $I(X;Y)$ between an input $X$ with probability mass function $\{\Lambda_{i,x}\}_{x\in\mathcal{X}}$ and the output produced by the channel (see property \eqref{JS-I} of the JS divergence), and thus is smaller than the capacity $C$.

\subsection{Proof of Proposition~\ref{Prop:EJSmax}}\label{app:EJSMAX}

Fix a time $t$ and assume that $Y^{t-1}=y^{t-1}$. 
Recall that $\Gamma^{\textnormal{PM}}$ denotes the random encoding function of the variable-length posterior matching scheme in Section~\ref{subsec:pm}.
By definition~\eqref{eq:gammamax} and by Proposition~\ref{prop:PM},
\begin{align}
\nonumber
EJS(\boldsymbol{\rho}(t),\gamma^*) &\geq EJS(\boldsymbol{\rho}(t),\Gamma^{\textnormal{PM}}) \geq C.
\end{align}

Now, assume that $\max_{i\in\Omega} \rho_i(t) \geq \tilde{\rho}$ and define 
\begin{equation}
\hat{i} := \argmax_{i\in\Omega} \rho_i(t).
\end{equation}
Then, 
\begin{equation} \label{eq:tilde}
\rho_{\hat{i}}(t)\ge \tilde{\rho}.
\end{equation}
Let $x,x' \in \mathcal{X}$ be two inputs of the channel satisfying
$D(P_x\|P_{x'})=C_1$. Also, define the encoding function
\begin{equation}
\hat{\gamma}(i): = \begin{cases}  x & \textnormal{ if } i =\hat{i} \\
 x' & \textnormal{ otherwise}.
 \end{cases}. 
 \end{equation}
%
By definition~\eqref{eq:gammamax}, from~\eqref{eq:tilde}, and by the selection of $x, x'$:
\begin{align}
EJS(\boldsymbol{\rho}(t),\gamma^*) &\ge EJS(\boldsymbol{\rho}(t),\hat{\gamma})
\ge \rho_{\hat{i}}(t) D(P_x\|P_{x'})
\ge \tilde{\rho} C_1.
\end{align}

\subsection{Proof of Proposition~\ref{Prop:DscrtH}}\label{app:proofHorstein}

Let 
\begin{equation}
\pi_{x}(t):=\sum_{i\in\Omega \colon \gamma^\mathrm{GHBZ}(i)=x} \rho_i(t), \qquad x\in\{0,1\}.
\end{equation}
Let
$$
k^*_2:=k^*- \text{sign} \bigg (\sum_{i=1}^{k^*} \rho_i(t)-\frac{1}{2} \bigg ),
$$
and define
$$
\delta_1(t):=\bigg|\sum_{i=1}^{k^*} \rho_i(t)-\frac{1}{2}\bigg|, \hspace*{.1in} \delta_2(t):=\bigg|\sum_{i=1}^{k^*_2} \rho_i(t)-\frac{1}{2}\bigg|. 
$$ 
Suppose $\sum_{i=1}^{k^*} \rho_i(t)-\frac{1}{2}<0$ which implies that $k^*_2=k^*+1$.
Note that by definition, $\pi_0(t)=\frac{1}{2}-\delta_1(t)$, $\rho_{k^*_2}(t)=\delta_1(t)+\delta_2(t)$, and $\pi_1(t)=\frac{1}{2}+\delta_1(t)$.
In this case, the EJS divergence is bounded as 
%
\begin{align*}
\lefteqn{EJS(\boldsymbol{\rho}(t),\gamma^\mathrm{GHBZ})}\\
&=\sum_{i=1}^{k^*} \rho_i(t) 
D\bigg(P_0 \Big\| \frac{\pi_0(t) - \rho_i(t)}{1-\rho_i(t)} P_0 + \frac{\pi_1(t)}{1-\rho_i(t)} P_1\bigg)\\
& \qquad + \rho_{k^*_2}(t) 
D\bigg(P_1 \Big\| \frac{\pi_0(t)}{1-\rho_{k^*_2}(t)} P_0 + \frac{\pi_1(t)-\rho_{k^*_2}(t)}{1-\rho_{k^*_2}(t)} P_1\bigg)\\
& \qquad + \sum_{i=k^*_2+1}^M \rho_i(t)
D\bigg(P_1 \Big\| \frac{\pi_0(t)}{1-\rho_i(t)} P_0 + \frac{\pi_1(t)-\rho_i(t)}{1-\rho_i(t)} P_1\bigg)\\
&\stackrel{(a)}{\ge} \pi_0(t) D\bigg(P_0 \Big\| \pi_0(t) P_0 + \pi_1(t) P_1\bigg)\\
& \qquad + \rho_{k^*_2}(t) D\bigg(P_1 \Big\| \frac{1}{2} P_0 + \frac{1}{2} P_1\bigg)\\ 
& \qquad + (\pi_1(t)-\rho_{k^*_2}(t)) D\bigg(P_1 \Big\| \pi_0(t) P_0 + \pi_1(t) P_1\bigg)\\
&\stackrel{(b)}{=} \pi_0(t) D\bigg(P_0 \Big\| \pi_0(t) P_0 + \pi_1(t) P_1\bigg)\\
& \qquad + \rho_{k^*_2}(t) D\bigg(P_0 \Big\| \frac{1}{2} P_0 + \frac{1}{2} P_1\bigg)\\ 
& \qquad + (\pi_1(t)-\rho_{k^*_2}(t)) D\bigg(P_0 \Big\| \pi_1(t) P_0 + \pi_0(t) P_1\bigg)\\
&\stackrel{(c)}{\ge} D\bigg(P_0 \Big\| \frac{1}{2} P_0 + \frac{1}{2} P_1\bigg)\\
&= C
\end{align*}
where $(a)$ follows from the facts that $\frac{\pi_0(t)-\rho_i(t)}{1-\rho_i(t)} \le \pi_0(t)$, $\frac{\pi_1(t)-\rho_{k^*_2}(t)}{1-\rho_{k^*_2}(t)} \le \frac{1}{2}$, $\frac{\pi_1(t)-\rho_i(t)}{1-\rho_i(t)} \le \pi_1(t)$, and by Lemma~\ref{DPQa};
$(b)$ holds because of condition \eqref{eq:channelcond};
and $(c)$ follows from the facts that KL divergence is convex, $(\pi_0(t))^2+\frac{1}{2}\rho_{k^*_2}(t)+(\pi_1(t)-\rho_{k^*_2}(t))\pi_1(t) =\frac{1}{2}+\delta_1(t)(\delta_1(t)-\delta_2(t))  \le \frac{1}{2}$, and by Lemma~\ref{DPQa}. 

The proof for the case $\sum_{i=1}^{k^*} \rho_i(t)-\frac{1}{2} \geq 0$ follows similarly. 

\subsection{Proof of Proposition~\ref{Prop:BSC}} \label{app:proofbinary}


Suppose $\gamma$ is an encoding function that satisfies \eqref{eq:detbinarycond}. 
Let $$\pi_{x}(t)=\sum\limits_{i\in\Omega \colon \gamma(i)=x} \rho_i(t) \ \ \ \text{for } x\in\mathcal{X}=\{0,1\},$$
and define $\delta(t)=\pi_0(t)-\pi_1(t)$.
From \eqref{eq:detbinarycond},
\begin{equation} 
0 \leq \delta(t) \leq \rho_i(t), \ \ \forall i \in \{j\in\Omega \colon \gamma(j)=0\}.
\end{equation}
 
We have
\begin{align*}
\lefteqn{EJS(\boldsymbol{\rho}(t),\gamma)}\\
&= \sum_{i=1}^M  \rho_i(t)
D\bigg(P_{\gamma(i)}\Big\|\sum_{j \neq i} \frac{\rho_j(t)}{1-\rho_i(t)} P_{\gamma(j)} \bigg) \\
&=\sum_{i\in\Omega \colon \gamma(i)=0} \rho_i(t) 
D\bigg(P_0 \Big\| \frac{\pi_0(t) - \rho_i(t)}{1-\rho_i(t)} P_0 + \frac{\pi_1(t)}{1-\rho_i(t)} P_1\bigg)\\
&\hspace*{.075in} + \sum_{i\in\Omega \colon \gamma(i)=1} \rho_i(t)
D\bigg(P_1 \Big\| \frac{\pi_0(t)}{1-\rho_i(t)} P_0 + \frac{\pi_1(t)-\rho_i(t)}{1-\rho_i(t)} P_1\bigg)\\
&\stackrel{(a)}{\ge} \sum_{i\in\Omega \colon \gamma(i)=0} \rho_i(t)
D\bigg(P_0 \Big\| \frac{1}{2} P_0 + \frac{1}{2} P_1\bigg)\\
&\hspace*{.075in} + \sum_{i\in\Omega \colon \gamma(i)=1} \rho_i(t)
D\bigg(P_1 \Big\| \frac{1}{2} P_0 + \frac{1}{2} P_1\bigg)\\
&\stackrel{(b)}{=} C
\end{align*}
where $(a)$ follows from the facts that $\pi_0(t)-\rho_i(t) \le \pi_1(t)$ 
for any $i$ with $\gamma(i)=0$, $\pi_1(t)\le \pi_0(t)$, and since
for two distributions $P$ and $Q$ and $\alpha\in[0,1]$, $D(P\|\alpha P + (1-\alpha) Q)$ is decreasing in $\alpha$ (see Lemma~\ref{DPQa});
and $(b)$ follows from Fact~\ref{Gallager} and since the capacity of the channel is achieved by the uniform input distribution.

On the other hand, if $\rho_{\hat{i}}(t)\ge\frac{1}{2}$, then condition~\eqref{eq:detbinarycond} is only satisfied by the encoding function $\hat{\gamma}$ 
under which $\hat{\gamma}(\hat{i})=0$ and $\hat{\gamma}(j)=1$ for all $j\neq \hat{i}$.
Therefore, if $\rho_{\hat{i}}(t)\ge \tilde{\rho}$ we obtain
\begin{align*}
EJS(\boldsymbol{\rho}(t),\hat{\gamma}) \ge \rho_{\hat{i}}(t) D(P_0\|P_1)
\ge \tilde{\rho} C_1.
\end{align*}

\subsection{Proof of Proposition~\ref{alg}}\label{app:new} 

For any encoding function $\gamma\in\mathcal{E}$, let
\begin{align}
\delta_{\gamma}(t)=\sum\limits_{i\in\Omega \colon \gamma(i)=0} \rho_i(t) - \sum\limits_{i\in\Omega \colon \gamma(i)=1} \rho_i(t).
\end{align}

Algorithm~\ref{AlgHM1} computes $\delta_{\gamma}(t)$ for all $2^M$ encoding functions $\gamma\in\mathcal{E}$
and selects $\gamma^\mathrm{Alg1}$ such that
\begin{align}
\label{Alg1cond}
\gamma^\mathrm{Alg1} := \argmin_{\gamma\in\mathcal{E} \colon \delta_{\gamma}(t)\ge 0} \delta_{\gamma}(t). 
\end{align}

Next we prove by contradiction that $\gamma^\mathrm{Alg1}$ satisfies \eqref{eq:detbinarycond}, i.e.,
\begin{equation} 
\delta_{\gamma^\mathrm{Alg1}}(t) \leq \rho_i(t), \ \ \forall i \in \{j\in\Omega \colon \gamma^\mathrm{Alg1}(j)=0\}.
\end{equation}
Suppose there exists $k\in\Omega$ such that $\gamma^\mathrm{Alg1}(k)=0$ and 
$\rho_k(t) < \delta_{\gamma^\mathrm{Alg1}}(t)$.
We consider two cases:

{\bf{Case I. $0 < \rho_k(t) \le \frac{1}{2} \delta_{\gamma^\mathrm{Alg1}}(t)$:}}

Define the encoding function $\hat{\gamma}_1$ as follows
\begin{equation}
\hat{\gamma}_1(i) = \begin{cases} 1 & \text{if } i=k\\ \gamma^\mathrm{Alg1}(i) & \text{otherwise} \end{cases}.
\end{equation} 
We have
\begin{align*}
0 \le \delta_{\hat{\gamma}_1}(t) = \delta_{\gamma^\mathrm{Alg1}}(t) - 2 \rho_k(t) < \delta_{\gamma^\mathrm{Alg1}}(t),
\end{align*}
which contradicts \eqref{Alg1cond}.

{\bf{Case II. $\frac{1}{2} \delta_{\gamma^\mathrm{Alg1}}(t) < \rho_k(t) < \delta_{\gamma^\mathrm{Alg1}}(t)$:}}

Define the encoding function $\hat{\gamma}_2$ as follows
\begin{equation}
\hat{\gamma}_2(i) = 1- \hat{\gamma}_1(i), \quad \forall i\in\Omega.
\end{equation} 
We have
\begin{align*}
0 < \delta_{\hat{\gamma}_2}(t) = 2 \rho_k(t) - \delta_{\gamma^\mathrm{Alg1}}(t) < \delta_{\gamma^\mathrm{Alg1}}(t),
\end{align*}
which again contradicts \eqref{Alg1cond}.


Algorithm~\ref{AlgHM2} constructs an encoding function that satisfies \eqref{eq:detbinarycond}.
Algorithm~\ref{AlgHM2} terminates in at most $M(M-1)/2$ rounds of operations, 
where in each round the main computational burden is to find an element of $S_0$ with the lowest belief.
Note that we do not have to search for the element with the lowest belief in each round if we sort all the beliefs once in the beginning, which has complexity order~$O(M \log M)$.

\bibliographystyle{IEEEtran}
\bibliography{HypTest} 

\begin{thebibliography}{10}
\providecommand{\url}[1]{#1}
\csname url@samestyle\endcsname
\providecommand{\newblock}{\relax}
\providecommand{\bibinfo}[2]{#2}
\providecommand{\BIBentrySTDinterwordspacing}{\spaceskip=0pt\relax}
\providecommand{\BIBentryALTinterwordstretchfactor}{4}
\providecommand{\BIBentryALTinterwordspacing}{\spaceskip=\fontdimen2\font plus
\BIBentryALTinterwordstretchfactor\fontdimen3\font minus
  \fontdimen4\font\relax}
\providecommand{\BIBforeignlanguage}[2]{{%
\expandafter\ifx\csname l@#1\endcsname\relax
\typeout{** WARNING: IEEEtran.bst: No hyphenation pattern has been}%
\typeout{** loaded for the language `#1'. Using the pattern for}%
\typeout{** the default language instead.}%
\else
\language=\csname l@#1\endcsname
\fi
#2}}
\providecommand{\BIBdecl}{\relax}
\BIBdecl

\bibitem{Burnashev76}
M.~V. Burnashev, ``{Data transmission over a discrete channel with feedback.
  Random transmission time},'' \emph{Problemy Peredachi Informatsii}, vol.~12,
  no.~4, pp. 10--30, 1975.

\bibitem{Yamamoto79}
H.~Yamamoto and K.~Itoh, ``{Asymptotic performance of a modified
  Schalkwijk--Barron scheme for channels with noiseless feedback},'' \emph{IEEE
  Transactions on Information Theory}, vol.~25, pp. 729--733, 1979.

\bibitem{Ooi98}
J.~M. Ooi and G.~W. Wornell, ``Fast iterative coding techniques for feedback
  channels,'' \emph{IEEE Transactions on Information Theory}, vol.~44, no.~7,
  pp. 2960--2976, November 1998.

\bibitem{Caire06}
G.~Caire, S.~Shamai, and S.~Verdu, ``Propagation, feedback and belief,''
  \emph{4th International Symposium on Turbo Codes \& Related Topics; 6th
  International ITG-Conference on Source and Channel Coding (TURBOCODING)}, pp.
  1--6, April 2006.

\bibitem{Tchamkerten06}
A.~Tchamkerten and E.~Telatar, ``Variable length coding over an unknown
  channel,'' \emph{IEEE Transactions on Information Theory}, vol.~52, pp.
  2126--2145, 2006.

\bibitem{Nakiboglu08}
B.~Nakiboglu and R.~G. Gallager, ``Error exponents for variable-length block
  codes with feedback and cost constraints,'' \emph{IEEE Transactions on
  Information Theory}, vol.~54, no.~3, pp. 945--963, March 2008.

\bibitem{Horstein63}
M.~Horstein, ``Sequential transmission using noiseless feedback,'' \emph{IEEE
  Transactions on Information Theory}, vol.~9, no.~3, pp. 136--143, July 1963.

\bibitem{Burnashev74}
M.~V. Burnashev and K.~S. Zigangirov, ``An interval estimation problem for
  controlled observations,'' \emph{Problemy Peredachi Informatsii}, vol.~10,
  no.~3, pp. 51--61, 1974.

\bibitem{Shayevitz11}
O.~Shayevitz and M.~Feder, ``Optimal feedback communication via posterior
  matching,'' \emph{IEEE Transactions on Information Theory}, vol.~57, no.~3,
  pp. 1186--1222, March 2011.

\bibitem{LiGamal14}
C.~T. Li and A.~E. Gamal, ``An efficient feedback coding scheme with low error
  probability for discrete memoryless channels,'' March 2014, submitted to
  \emph{IEEE Transactions on Information Theory}.

\bibitem{Lin91}
J.~Lin, ``{Divergence measures based on the Shannon entropy},'' \emph{IEEE
  Transactions on Information Theory}, vol.~37, no.~1, pp. 145--151, January
  1991.

\bibitem{toddISIT}
T.~P. Coleman, ``A stochastic control viewpoint on `posterior matching'-style
  feedback communication schemes,'' in \emph{IEEE International Symposium on
  Information Theory (ISIT)}, 2009, pp. 1520--1524.

\bibitem{Jeffreys46}
H.~Jeffreys, ``An invariant form for the prior probability in estimation
  problems,'' \emph{Proceedings of the Royal Society. London. Series A.}, vol.
  186, pp. 453--461, 1946.

\bibitem{Burbea82b}
J.~Burbea and C.~R. Rao, ``On the convexity of some divergence measures based
  on entropy functions,'' \emph{IEEE Transactions on Information Theory},
  vol.~28, no.~3, pp. 489--495, May 1982.

\bibitem{Toussaint71}
G.~T. Toussaint, ``Some functional lower bounds on the expected divergence for
  multihypothesis pattern recognition, communication, and radar systems,''
  \emph{IEEE Transactions on Systems, Man, and Cybernetics}, vol. SMC-1, pp.
  384--385, 1971.

\bibitem{Gastpar2010}
M.~C. Gastpar, P.~R. Gill, A.~G. Huth, and F.~E. Theunissen, ``Anthropic
  correction of information estimates and its application to neural coding,''
  \emph{IEEE Transactions on Information Theory}, vol.~56, no.~2, pp. 890--900,
  February 2010.

\bibitem{CoverBook}
T.~M. Cover and J.~A. Thomas, \emph{Elements of information theory},
  2nd~ed.\hskip 1em plus 0.5em minus 0.4em\relax Hoboken, NJ: John Wiley \&
  Sons, Inc., 2006.

\bibitem{Gallager68}
R.~G. Gallager, \emph{Information theory and reliable communication}.\hskip 1em
  plus 0.5em minus 0.4em\relax New York, NY: John Wiley \& Sons, Inc., 1968.

\bibitem{Polyanskiy11}
Y.~Polyanskiy, H.~V. Poor, and S.~Verdu, ``Feedback in the non-asymptotic
  regime,'' \emph{IEEE Transactions on Information Theory}, vol.~57, no.~8, pp.
  4903--4925, August 2011.

\bibitem{Berlin09}
P.~Berlin, B.~Nakiboglu, B.~Rimoldi, and E.~Telatar, ``{A simple converse of
  Burnashev's reliability function},'' \emph{IEEE Transactions on Information
  Theory}, vol.~55, pp. 3074--3080, 2009.

\bibitem{Witsenhausen1968}
H.~S. Witsenhausen, ``A counterexample in stochastic optimum control,''
  \emph{SIAM Journal on Control}, vol.~6, no.~1, pp. 131--147, 1968.

\bibitem{Mahajan08}
A.~Mahajan, A.~Nayyar, and D.~Teneketzis, ``Identifying tractable decentralized
  problems on the basis of information structures,'' in \emph{46th Annual
  Allerton Conference on Communication, Control, and Computing}, 2008, pp.
  1440--1449.

\bibitem{HypJournal}
M.~Naghshvar and T.~Javidi, ``Active sequential hypothesis testing,'' \emph{The
  Annals of Statistics}, vol.~41, no.~6, pp. 2703--2738, 2013.

\bibitem{Bertsekas}
D.~P. Bertsekas and J.~N. Tsitsiklis, \emph{Neuro-Dynamic Programming}.\hskip
  1em plus 0.5em minus 0.4em\relax Belmont, MA: Athena Scientific, 1996.

\bibitem{DeGroot70}
M.~H. DeGroot, \emph{{Optimal Statistical Decisions}}.\hskip 1em plus 0.5em
  minus 0.4em\relax New York, NY: McGraw-Hill Book Co., 1970.

\bibitem{ISIT2012}
M.~Naghshvar and T.~Javidi, ``{Extrinsic Jensen--Shannon divergence with
  application in active hypothesis testing},'' in \emph{IEEE International
  Symposium on Information Theory Proceedings (ISIT)}, July 2012, pp.
  2191--2195.

\bibitem{ITW2012}
M.~Naghshvar, M.~Wigger, and T.~Javidi, ``Optimal reliability over a class of
  binary-input channels with feedback,'' in \emph{IEEE Information Theory
  Workshop (ITW)}, September 2012, pp. 391--395.

\bibitem{Burnashev75}
M.~V. Burnashev and K.~S. Zigangirov, ``On one problem of observation
  control,'' \emph{Problemy Peredachi Informatsii}, vol.~11, no.~3, pp. 44--52,
  1975.

\end{thebibliography}

\balance

\begin{IEEEbiographynophoto}{Mohammad Naghshvar}
(S'07-M'13) received the B.S. degree in electrical engineering from Sharif University of Technology in 2007. 
He obtained the M.Sc. degree and the Ph.D. degree in electrical engineering (communication theory and systems)
both from University of California San Diego in 2009 and 2013, respectively. 
He is currently a senior R\&D engineer at Qualcomm Technologies Inc., San Diego, CA.  
His research interests include active learning and hypothesis testing, 
stochastic control and optimization, wireless communication and information theory.
\end{IEEEbiographynophoto}

\begin{IEEEbiographynophoto}{Tara Javidi}
(S'96-M'02-SM'12) studied electrical engineering at the Sharif University of Technology from 1992 to 1996. She received her MS degrees in Electrical Engineering: Systems, and Applied Mathematics: Stochastics, from the University of Michigan, Ann Arbor, MI. She received her PhD 
in electrical engineering and computer science from the University of Michigan, Ann Arbor, in 2002.
From 2002 to 2004, she was an Assistant Professor at the Electrical Engineering Department, University of Washington, Seattle. In 2005, she joined University of California, San Diego, where she is currently an Associate Professor of electrical and computer engineering.

Tara Javidi was a Barbour Scholar during 1999-2000 academic year and received an NSF CAREER Award in 2004. Her research interests are in communication networks, stochastic resource allocation, stochastic control theory, and wireless communications.
\end{IEEEbiographynophoto}

\begin{IEEEbiographynophoto}{Mich\`ele Wigger}
(S'05-M'09-SM'14) received the M.Sc. degree in electrical
engineering (with distinction) and the Ph.D. degree in electrical engineering
both from ETH Zurich in 2003 and 2008, respectively. In 2009 she was
a postdoctoral researcher at the ITA center at the University of California,
San Diego. Since December 2009 she is an Assistant Professor at Telecom
ParisTech, in Paris, France. Her research interests are in information and
communications theory; in particular in wireless networks, feedback channels,
 channels with states, and distributed source coding.
\end{IEEEbiographynophoto}

\end{document}